\documentclass[a4paper, 11pt]{article}
\usepackage{graphicx}
\usepackage{epsf,amsmath,bbold,amsfonts,stmaryrd}

\usepackage[utf8]{inputenc}
\usepackage{mathrsfs}
\usepackage{appendix}
\usepackage{amssymb}
\usepackage{float}
\usepackage{color}
\usepackage{cite}
\usepackage{hyperref}
\hypersetup{pageanchor=false}
\usepackage{indentfirst}
\usepackage{url}
\usepackage{float}
\usepackage{caption}
\usepackage[numbers,square,comma,sort&compress,merge]{natbib}

\usepackage{subfigure}

\usepackage{ulem}

\def\nn{\nonumber}

\def\be{\begin{equation}}
\def\ee{\end{equation}}

\def\bea{\begin{eqnarray}}
\def\eea{\end{eqnarray}}

\def\ba{\begin{array}}
\def\ea{\end{array}}

\def\bc{\begin{center}}
\def\ec{\end{center}}

\def\bl{\begin{flushleft}}
\def\el{\end{flushleft}}

\def\br{\begin{flushright}}
\def\er{\end{flushright}}

\def\bi{\begin{itemize}}
\def\ei{\end{itemize}}

\def\bt{\begin{tabular}}
\def\et{\end{tabular}}

\newcommand{\N}{N_{\text{max}}}

\newcommand{\con}{\text{constant}}
\newcommand{\xc}{x_\text{c}}
\newcommand{\yc}{y_\text{c}}
\newcommand{\xmi}{x_\text{min}}
\newcommand{\xma}{x_\text{max}}
\newcommand{\ymi}{y_\text{min}}
\newcommand{\yma}{y_\text{max}}

\numberwithin{equation}{section}

\begin{document}
\title{\textbf{Shadow of Kerr black hole with an Gaussian distributed plasma in the polar direction}}

\author{Zhenyu Zhang$^{1}$,  Haopeng Yan$^2$, Minyong Guo$^{3*}$ and Bin Chen$^{1,2,4}$}
\date{}
\maketitle

\vspace{-10mm}

\begin{center}
{\it
$^1$School of Physics, Peking University, No.5 Yiheyuan Rd, Beijing
100871, P.R. China\\\vspace{1mm}

$^2$Center for High Energy Physics, Peking University,
No.5 Yiheyuan Rd, Beijing 100871, P. R. China\\\vspace{1mm}

$^3$Department of Physics, Beijing Normal University, Beijing 100875, P. R. China\\\vspace{1mm}

$^4$Collaborative Innovation Center of Quantum Matter, No.5 Yiheyuan Rd, Beijing 100871, P. R.
China
}
\end{center}

\vspace{8mm}

\begin{abstract}
In this work, we study the shadow of Kerr black hole surrounded by an axisymmetric plasma, whose density takes an Gaussian distribution in the polar direction. Along the radial direction, we consider two models: in model A  the density of the plasma decays in a power law; in model B the density obeys a logarithmic Gaussian distribution. Using the numerical backward ray-tracing method, we find that the size of the shadow is sensitive to the inclination angle of the observer due to the polar distribution of the density of the plasma. In particular, we pay special attention to the model B and investigate the influence of the radial position of maximum density, the decay rate of the density towards the event horizon and the opening angle of the plasma on the shape and size of the Kerr black hole shadow. The effects of  the plasmas studied in this work  can be qualitatively explained by taking the plasmas as convex lenses with the refractive index being less than $1$. 

\end{abstract}

\vfill{\footnotesize $^*$Corresponding author: minyongguo@bnu.edu.cn.}

\maketitle

\newpage

\section{Introduction}

Black hole is one of the most fascinating objects in general relativity. The direct evidences for the existence of black holes have  been accumulated  in the past few  years.  In particular, the Event Horizon Telescope (EHT) Collaborate released the images of M87* three years ago \cite{EventHorizonTelescope:2019dse} and the image of Sgr A*   very recently \cite{EventHorizonTelescope:2022xnr}.   As we know, black holes are black because they swallow any matter falling into the horizons and not even light can escape, so that we cannot actually see a black hole in any direct way. Thus, the images of black holes  indicate that  the supermassive black holes in the centers of galaxies are not isolated, and they are actually surrounded by plasma, corona and magnetic field configuration. Among these surroundings, the plasma is the most important one,  and  plays an indispensable role in forming the pictures of black holes. 

On the one hand, the accretion disk of plasma can be used as a light source to illuminate the black hole. On the other hand, plasma as a dispersive medium would affect the path of the light traveling through it. In this present work, we would like to focus on the latter. Along this line, the null equation of motion including the influence of a non-magnetized pressureless plasma in the Schwarzschild spacetime was studied in \cite{synge1960relativity, 2000Ray, Rogers:2015dla, Bisnovatyi-Kogan:2008qbk, Atamurotov:2021hoq}. The gravitational lensing and light deflection by other static black holes and rotational black holes in the same class of plasma medium can be found in \cite{Bisnovatyi-Kogan:2010flt, Bisnovatyi-Kogan:2017kii, Benavides-Gallego:2018ufb, Tsupko:2013cqa, MAT, Er:2013efa, Liu:2016eju, Fathi:2021mjc, Babar:2021nst, Matsuno:2020kju, Jin:2020emq}. The black hole shadow in the presence of a plasma are investigated in \cite{Perlick:2015vta, Perlick:2017fio, Abdujabbarov:2015pqp, Atamurotov:2015nra, Babar:2020txt, Ahmedov:2019dja, Atamurotov:2021cgh, Yan:2019etp, Wang:2021irh, Badia:2021kpk, Li:2021btf, Chowdhuri:2020ipb, Kala:2022uog}. In these works, either the electron frequency of a non-magnetic cold plasma are assumed to take a special form to make the Hamilton-Jacobi equation of photons separable for rotating black hole spacetimes, or only weak gravitational lensings are considered, such that analytical methods can be employed. However, in order to gain more understandings of the influence of plasma on black hole shadow, more realistic models should be studied. In \cite{Huang:2018rfn}, the authors considered the models in which the radial density of the plasma takes a power-law or an exponential form, and calculated the Kerr black hole shadow using the numerical backward ray-tracing method. In their study,  the polar distribution of the plasma was assumed to be still uniform, that is, the plasma is spherically symmetric.

For a realistic model, the plasma cannot be spherically symmetric around a spinning black hole. 
And for the radial density, it is expected to be mostly not monotonically decaying,   instead 
it reaches maximum at some point and decay toward the event horizon and the infinity. It was supposed that the rate of decay toward the horizon is  faster than that toward the infinity\cite{abramowicz1978relativistic,kozlowski1978analytic}. In this work, we consider two models of plasma, in both of which the polar density  takes an Gaussian distribution, and the radial density takes the power-law form in one model  and  a logarithmic Gaussian form in the other. In particular, we discuss the effect of plasma on the black hole shadow by showing that the plasma has a convex lens effect for the light rays.

The remaining parts of this paper are organized as follows. In section \ref{section2}, we introduce the background spacetime and the plasma models to set up our problems. In section \ref{sec3}, we show our numerical method, present and discuss the results. The main conclusions are summarized in section \ref{summary}. In this work, we have set the fundamental constants $c$, $G$, the vacuum permittivity $\varepsilon_0$ and the mass of the black hole $M$ to unity, and we will work in the signature convention $(-, +, +, +)$ for the spacetime metric.

\section{Kerr black hole and plasma models}\label{section2}
In this section, we would like to set up our problem. The background of interest is  a Kerr spacetime,  whose metric takes a form 
\bea
ds^2=-\left(1-\frac{2r}{\Sigma}\right)dt^2+\frac{\Sigma}{\Delta}dr^2+\Sigma d\theta^2+\frac{1}{\Sigma}\left[(r^2+a^2)^2-\Delta a^2\sin^2\theta \right]\sin^2\theta d\phi^2-\frac{4ar}{\Sigma}\sin^2\theta dtd\phi\nn\\\,,
\eea
in the Boyer-Lindquist coordinates, where 
\bea
\Delta=r^2-2r+a^2\,,\quad\quad\Sigma=r^2+a^2\cos^2\theta.
\eea
Note that, we have set $M=1$ for simplicity and without loss of generality. The black hole horizon is located at $r_h=1+\sqrt{1-a^2}$ which is the larger root of the equation $\Delta=0$.
We assume that outside the black hole there exist some refracting medium which have no backreaction to the background \footnote{In the following, we assume that the refracting medium is cold plasma. In Appendix. \ref{appA}, we estimate the numerical values of the plasma effects on the background spacetime and confirm that the self-gravity effect of the plasma considered in our work can be neglected.}. In the presence of the refracting medium, the Hamiltonian of photons is given by\cite{synge1960relativity} 
\bea\label{ham}
H=\frac{1}{2}\left[g_{\mu\nu}p^\mu p^\nu-(n^2-1)(p_\mu V^\mu)^2\right]\,,
\eea
where, $n$ is the refractive index and $V^\mu$ is defined to denote the $4$-velocity of the refracting medium. Considering the refracting medium is composed of a non-magnetized pressureless plasma, we have 
\bea\label{nsq}
n^2=1-\frac{\omega_p^2}{\omega^2}.
\eea
Here $\omega=-p_\mu V^\mu$ is the frequency of photon and $\omega_p$ is the plasma electron frequency satisfying 
\bea\label{omegap}
\omega_p^2=\frac{4\pi e^2}{m_e}N\,,
\eea
where $e$ and $m_e$ are the charge and mass of the electron, respectively, and $N$ is the number density of the plasma which is generally a function of the spacetime coordinates $x^\mu$. In addition, considering the Kerr spacetime is stationary and axisymmetric, it is plausible to assume the number density $N$ is independent of the coordinates $t$ and $\phi$, that is, the plasma is distributed radially and angularly. In this work, in the polar direction, we take normal distribution  for simplicity. As for the radial direction, we would like to consider two models. One is that the plasma decays as a power law along the radial direction and the other is set to obey logarithmic normal distribution. Precisely, for the first model A, the electron frequency of the plasma takes the form
\bea\label{omeA}
\omega_{pA}^2(r,\theta)=\frac{k_A}{r^2}e^{-\frac{(\theta-\pi/2)^2}{2\xi_\theta^2}}\,, \hspace{3ex}r>r_h,
\eea
and correspondingly, we have the number density 
\bea
N_A(r,\theta)=\N \frac{r_h^2}{r^2}e^{-\frac{(\theta-\pi/2)^2}{2\xi_\theta^2}}\,,\hspace{3ex}r>r_h,
\eea
where $k_A$ is a certain constant number to characterize the number density, $\N=\frac{k_Am_e}{4\pi e^2r_h^2}$ is a rescaled parameter of $k_A$, being equal to the number density near the black hole horizon in this case, and $\xi_\theta$ is defined as the shape parameter to characterize the dispersion degree of the normal distribution. Note that in \cite{Huang:2018rfn}, the authors assumed $\omega_{p}^2\propto r^{-h}$, where $h$ is a positive number, and studied the influence of the value of $h$ on the Kerr shadow. In this work, we would like to focus on the effect of the distribution in the polar direction, thus we just fix the radial dependence to be $1/r^2$ for simplicity.  According to the normal distribution, we can see that the plasma is concentrated in the region $[\pi/2-2\xi_\theta, \pi/2+2\xi_\theta]$ when $\xi_\theta$ is less than $\pi/4\approx0.785$, and thus $4\xi_\theta$ can be approximatively taken as the opening angle of the plasma accretion. However, when $\xi_\theta$ is ver large, the polar normal distribution is approximately uniform, and our model reduce to one of the models studied in \cite{Huang:2018rfn}. This allows us to check our calculations with the one in \cite{Huang:2018rfn} by taking the large $\xi_\theta$ limit. 

\begin{figure}[h!]
  \centering
  \includegraphics[width=3.5in]{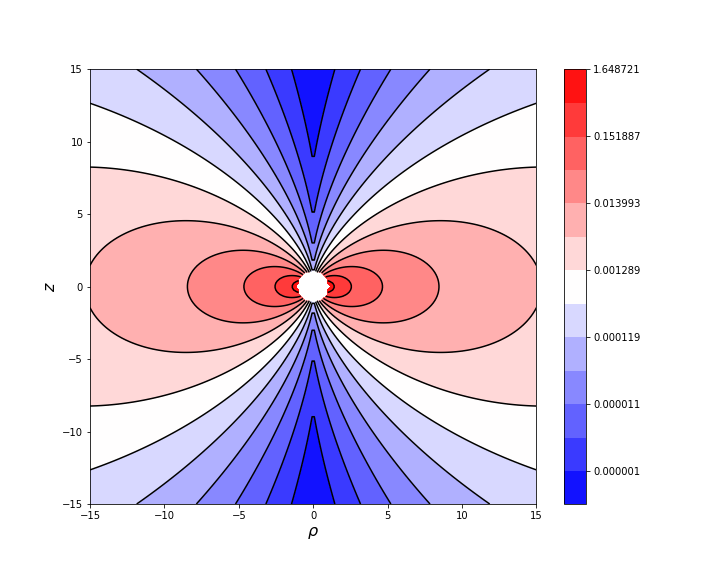}
%  \subfigure[$$]{
%  \begin{minipage}[t]{0.45\linewidth}
%  \centering
%  \includegraphics[width=3.5in]{model_A_2d.png}
%  \end{minipage}
%  }
  
  \centering
  \caption{Density contour map of a two dimensional slice $\phi=\con$ for model A of the plasma with $\N=1$ and $\xi_\theta=\pi/9$. The horizontal and vertical coordinates are $\rho=r\sin\theta$ and $z=r\cos\theta$ respectively, where $\theta$ and $\phi$ are polar and azimuthal angles in the Boyer-Lindquist coordinate.} 
  \label{modelA}
\end{figure}

\begin{figure}[h!]
  \centering
  \includegraphics[width=3.5in]{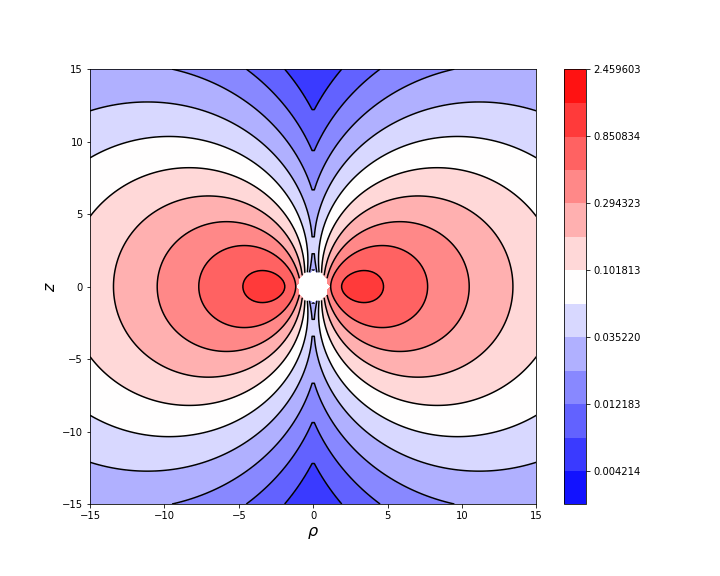}
%  \subfigure[$$]{
%  \begin{minipage}[t]{0.45\linewidth}
%  \centering
%  \includegraphics[width=3.5in]{model_B_2d.png}
%  \end{minipage}
%  }
  \centering
  \caption{Density plots of the plasma for model B. The coordinates are the same with those in Fig. \ref{modelA}. $\N=1$, $\sigma=0.8$, $r_m=3$ and $\xi_\theta=\pi/9$. }
  \label{modelB}
\end{figure}

In Fig. \ref{modelA}, we show the density plots of the plasma for model A with $\N=1$ and $\xi_\theta=\pi/9$. It is not hard to see that in this model, the plasma is mainly centered in the vicinity of the Kerr black hole horizon. Hence, although this model is able to capture some characteristics of the plasma around a black hole, such as the refractive index, rapid decay at large scales, etc., there are a few disadvantages that cannot be ignored. One of the most obvious shortcoming is that the density should not increase monotonically as the plasma gets closer to the event horizon. It is believed that the density reach a maximum at a certain radius $r_m$ and fall off to the event horizon as well as to the infinity, with the rate of decay towards the event horizon being larger \cite{abramowicz1978relativistic,kozlowski1978analytic}. To capture this important feature of the plasma, we introduce the second model, which will be referred to as model B and has a lognormal distribution along the radial direction. More precisely, we have 
\bea\label{omeB}
\omega_{pB}^2(r, \theta)=k_Be^{-\frac{\left(\log\frac{r}{r_m}\right)^2}{2\sigma^2}}e^{-\frac{(\theta-\pi/2)^2}{2\xi_\theta^2}}\,, \hspace{3ex}r>r_h,
\eea
and  the corresponding number density of the plasma 
\bea
N_B(r, \theta)=\N e^{-\frac{\left(\log\frac{r}{r_m}\right)^2}{2\sigma^2}}e^{-\frac{(\theta-\pi/2)^2}{2\xi_\theta^2}}\,,\hspace{3ex}r>r_h,
\eea
where $r_m$ is the the position of the maximum density, $\N=N_B(r_m,\pi/2)$ and $\sigma$ is one of the parameters of lognormal distribution. We want to stress that our lognormal distribution is slightly different from the most common form, the reason is that we have taken into account of  the normalization. In Fig. \ref{modelB}, we show the density plots of the plasma for model B with $\N=1$, $\sigma=0.8$, $r_m=3$ and $\xi_\theta=\pi/9$. From this figure, we can see that the position with the maximum density is no longer at the event horizon and the density declines faster towards the event horizon than to the infinity.  These results are in line with our expectations for the plasma.

\begin{figure}[h!]

  \centering
  \includegraphics[width=3.5in]{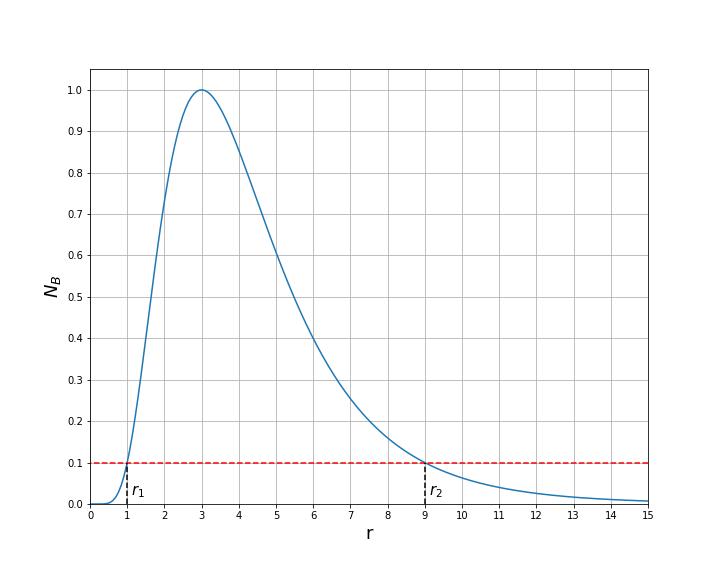}

  \caption{A profile of a lognormal distribution with $r_m=3$ and $\sigma=0.512$. The width of the accretion is $d=8$.}
  \label{lognormal}
\end{figure}

In addition, the density profile of our plasma model B behaves qualitatively like a true accretion disk. Therefore, we may take the plasma model B as a plasma accretion, of which the number density $N_B$ is above $0.1\N$. Then, we would like to introduce a parameter $d$ to manifest the width of the plasma accretion
\bea\label{dr}
d=r_2-r_1\,,
\eea
where $r_{1,2}$ are the cutoff positions of the plasma with
\be
r_{1,2}=r_m e^{\pm\sqrt{2\log10}\sigma}\ee
 being the roots of the equation $e^{-\frac{\left(\log\frac{r}{r_m}\right)^2}{2\sigma^2}}=0.1$.   We give an example shown in Fig. \ref{lognormal}, where we choose $r_m=3$ and $d=8$ with $r_1=1$ and $r_2=9$. %From Fig. \ref{lognormal}, it is also easy to see that $r_m-r_1$ is smaller than $r_2-r_m$, which once again shows that the dropoff rate of the number density is smaller on the side $r>r_m$.

Now we have completed the introduction of the spacetime background and the plasma models. After plunging Eq. (\ref{nsq}) into Eq. (\ref{ham}), we obtain the Hamiltonian 
\bea\label{hamf}
H(x^\mu, p_\mu)=\frac{1}{2}(g_{\mu\nu}p^\mu p^\nu+\omega^2_p)\,,
\eea
where $\omega_p^2$ are given by the Eq. (\ref{omeA}) for model A and Eq. (\ref{omeB}) for  model B. From this Hamiltonian we can study the motions of the photons in the backgrounds. 

\section{Black hole shadows}\label{sec3}

In this section, we are going to study the shadows of the Kerr black holes surrounded by a plasma  with the distribution models mentioned in sec. \ref{section2}. Considering the Eqs. (\ref{omeA}), (\ref{omeB}) and  (\ref{hamf}), we can see that the Hamiltonian including $\omega_{pA}^2$ or $\omega_{pB}^2$ is no longer separable, and the standard analytical method to calculate black hole shadows can not apply here. Therefore, we would like to apply the numerical backward ray-tracing method  proposed in \cite{Hu:2020usx, Zhong:2021mty} for our study. As our interest is focused on the influence of plasma in this work and the Kerr shadow deviates more from the circle with a larger spin, we fix $a=0.998$ in the following discussion, considering astrophysical black holes have the maximum spin $a=0.998$ which is known as the Thorne's limit \cite{Thorne:1974ve}.

It is very convenient to do the numerical geodesic evolution by using the first-order differential equations. In the Hamiltonian canonical formalism, the geodesic equations read
\bea\label{hcf}
\dot{p}_\mu=-\frac{\partial H}{\partial x^\mu}\,,\quad\quad \dot{x}^\mu=\frac{\partial H}{\partial p_\mu}
\eea
 where the dot denotes the derivative with respect to the affine parameter $\tau$. Recall that $\omega_p^2$ is independent on $t$ and $\phi$, we have two conserved quantities
\bea
E=-p_t\,,\quad\quad L=p_\phi\,,
\eea
along the null geodesics. In the local rest frame of the observer at the position $(t_o, r_0, \theta_o, \phi_o)$, one can use celestial coordinates to denote the $3$-momentum vector of $p^\mu$, and find the timelike component after considering the Hamiltonian $H=0$. In addition, on the screen of the observer, one can also set up standard Cartesian coordinates and then build a map from the celestial coordinates and the Cartesian coordinates. Here, we also employ the stereographic projection, which is often called fisheye camera model. The details of this model can be found in \cite{Hu:2020usx}. Hereto, we have known the values of $(x^\mu, p_\mu)$ for given Cartesian coordinates on the screen of the observer, which can be set as the initial values of the Eq. (\ref{hcf}), then we can determine the trajectories of the photons and identify whether the rays fall into the black hole or reach the infinity. To fix the black hole shadow, our strategy is to place a spherical source illuminating the system, that is, the black hole and the observer are both inside the spherical source. As a result, the pixels on the screen of the observer will be coloured if the corresponding rays can reach the source, otherwise, photons fall into the black hole and the pixels are dark. The boundary between the dark and coloured region is the shadow curve we want.

%\subsection{Results}\label{result}

\begin{figure}[h!]

  \centering
  \includegraphics[width=3.5in]{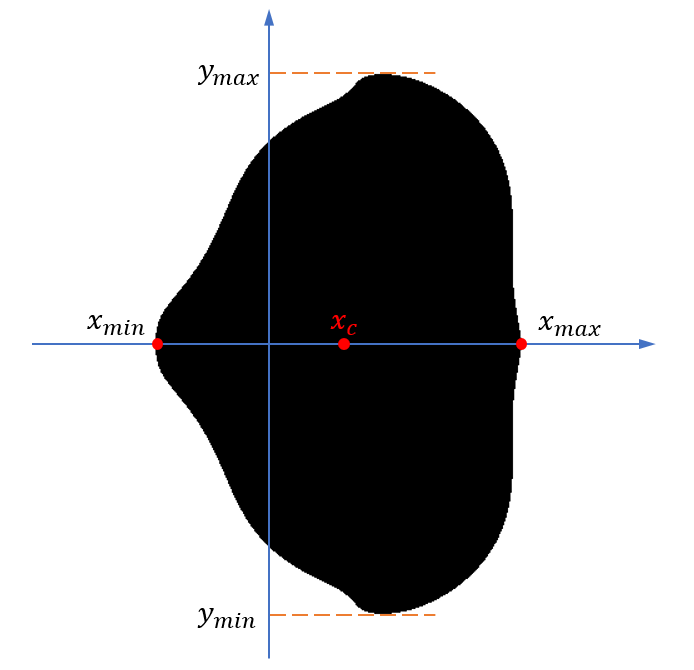}

  \caption{A diagram of the Kerr black hole shadow curve, deformed by the plasma around the black hole. The shadow in this diagram is given with model B and $k_B=0.8$, $r_m=5$, $\sigma=0.18$.}
  \label{cd}
\end{figure}

%In this subsection, we are going to give the results of Kerr black hole shadows in the presence of Gaussian distributed plasma. Note that 

In order to quantitatively express the deformation of the shadow of a vacuum Kerr black hole and the variation of the shape of the shadow of a Kerr black hole surrounded with the plasma in different distributions, we introduce the parameter $\bar{R}$ to reflect the feature of the black hole shadow size. In Fig. \ref{cd}, we  show the basic coordinate parameters on the screen of the observer. The origin of the coordinates is the stereographic projection of the observer on the screen. The parameters $\xmi$ and $\xma$ are the minimum and maximum of the shadow curve on the horizontal axis, respectively. Similarly, we have $\ymi$ and $\yma$ along the vertical direction. It is worth noting that $\ymi=-\yma$ due to the $\mathcal{Z}_2$ symmetry of the spacetime. Then we can define the centre of the shadow to be $(\xc, \yc)$ with $\xc=\left(\xma+\xmi\right)/2$ and $\yc=0$. It is also convenient to introduce polar coordinates $(R, \psi)$ on the screen and the origin is placed at the centre $(\xc, \yc)$ such that $R=\sqrt{\left(x-\xc\right)^2+\yc^2}$. Then we define the parameter $\bar{R}$ as 
\bea
\bar{R}=\int_0^{2\pi}\frac{R(\psi)}{2\pi}d\psi. 
\eea
It represents the average radius of the shadow curve, characterizing the size of the black hole shadow. 

\subsection{Model A}

\begin{figure}[h!]
  \centering
  \subfigure[$\theta_o=\pi/2, \xi_\theta=10$]{
  \begin{minipage}[t]{0.3\linewidth}
  \centering
  \includegraphics[width=1.5in]{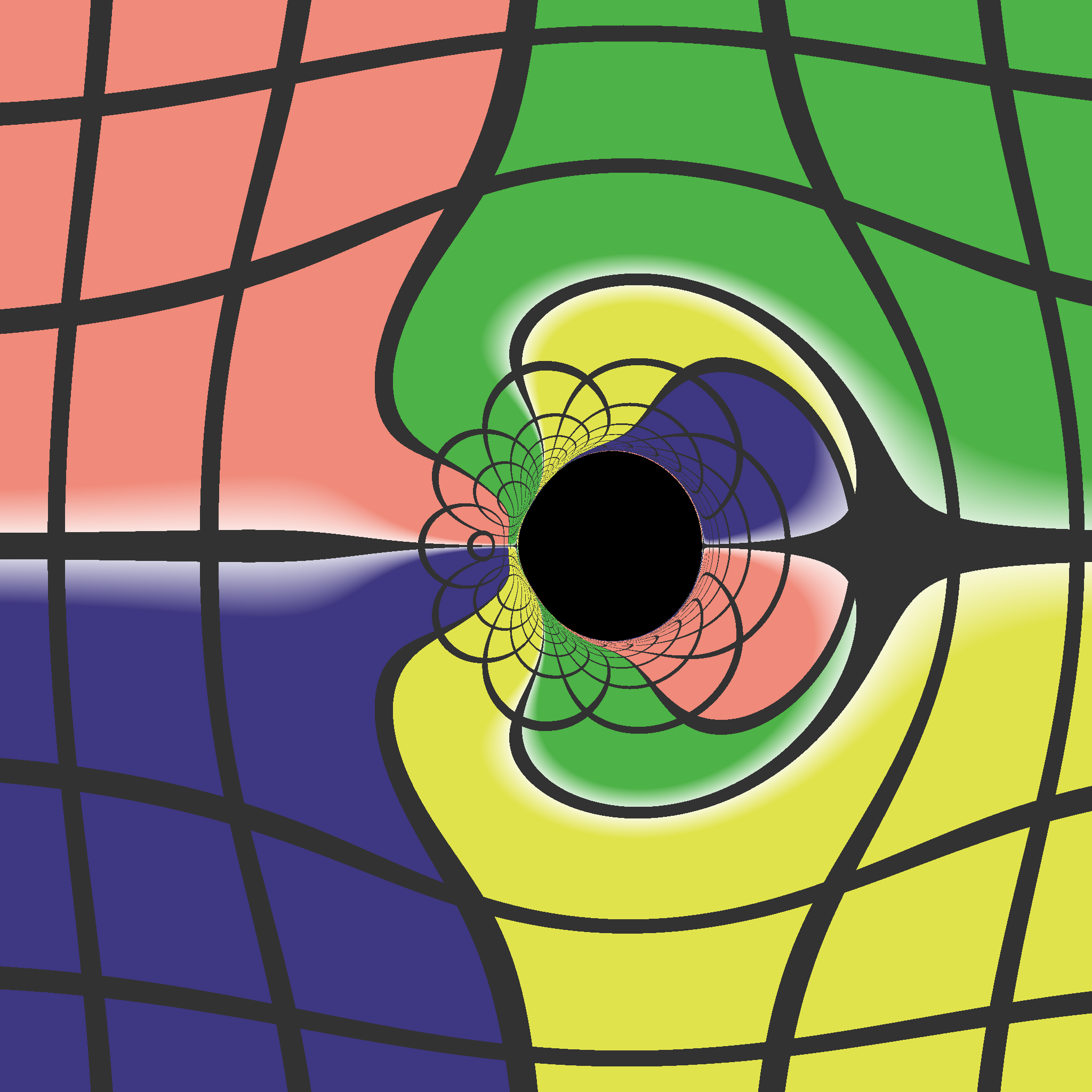}
  \end{minipage}
  }%
  \subfigure[$\theta_o=\pi/2, \xi_\theta=0.09$]{
  \begin{minipage}[t]{0.3\linewidth}
  \centering
  \includegraphics[width=1.5in]{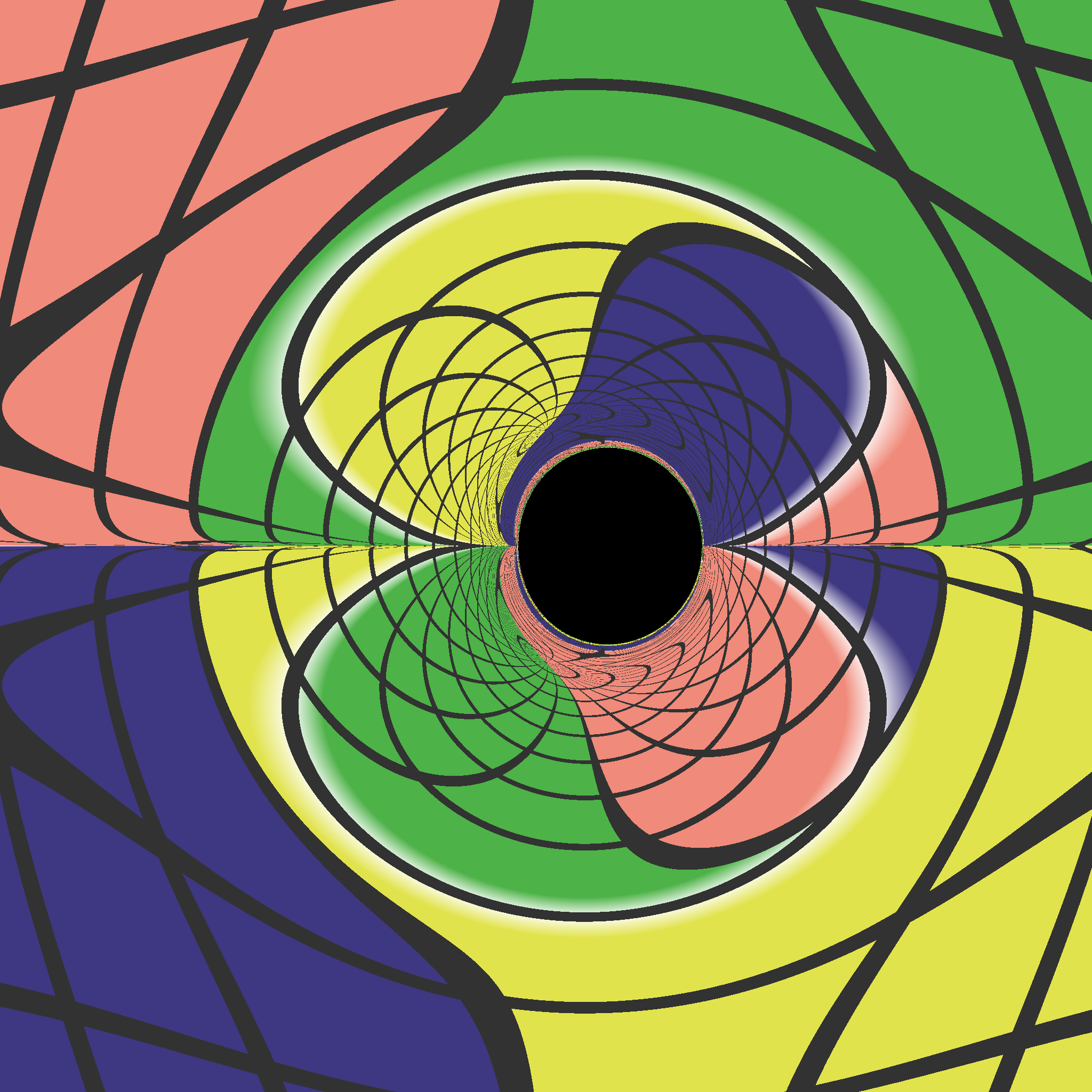}
  \end{minipage}%
  }%
  \subfigure[$\theta_o=17^\circ, \xi_\theta=0.18$]{
  \begin{minipage}[t]{0.3\linewidth}
  \centering
  \includegraphics[width=1.5in]{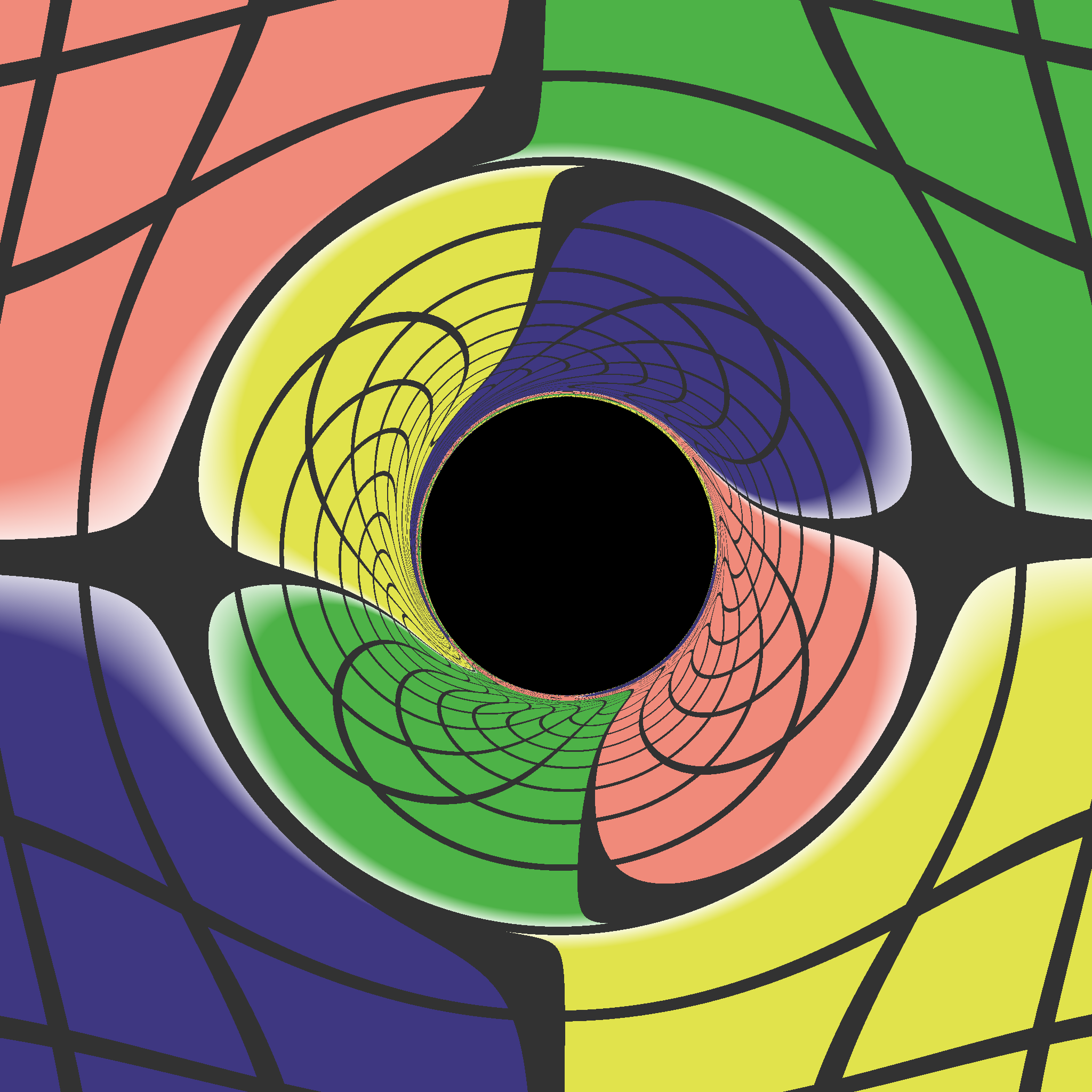}
  \end{minipage}%
  }%
  \centering
  \caption{The images of the Kerr black hole surrounded by a plasma of model A. The spin is fixed at $a=0.998$ and $k_A=16$.}
  \label{mAxi}
\end{figure}

Firstly, let us show the results for model A. In Fig. \ref{mAxi}, we set $k_A=16$ and present some selected images of the Kerr black hole. For the first two plots of Fig. \ref{mAxi}, we take $\theta_o=\pi/2$, which corresponds to the observer being located on the equatorial plane. And for the third plot, we set $\theta_o=17^\circ$ which is same with the inclination angle for the supermassive black hole in M87. In the first plot of  Fig. \ref{mAxi}, we set $\xi_\theta=10\gg\pi/4$ so that the distribution of plasma is approximately uniform in the polar direction and thus can be taken as a spherically symmetric plasma. Comparing the shape of the shadow and the gravitational lensings in this image with those in the last plot of Fig. 7 in \cite{Huang:2018rfn}, we find that they agree well, although we do not quantitatively compare the size of the shadows. In the last two plots in Fig. \ref{mAxi}, we set $\xi_\theta=0.09$ and $\xi_\theta=0.18$, respectively, which means that the plasma is no longer spherically symmetric and has a finite opening angle. In these two cases, it is not hard to see in Fig. \ref{mAxi} that the shapes of the shadow curves look similar, while the left side of the shadow curve in plot (b) gets sharp visibly, that is, the influence on the shape of the Kerr shadow curve are distinctly different for spherically symmetric and axisymmetric plasma.

\begin{figure}[h!]
  \centering

  \subfigure[$$]{
  \begin{minipage}[t]{0.45\linewidth}
  \centering
  \includegraphics[width=3.1in]{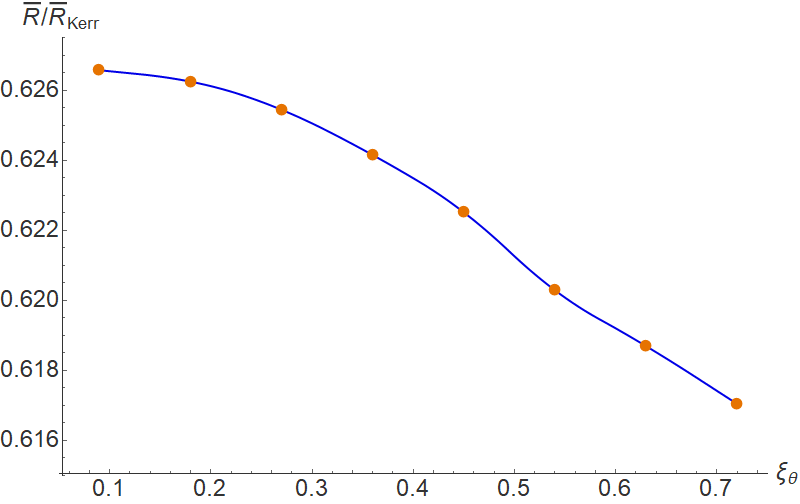}
  \end{minipage}
  }
  \subfigure[$$]{
  \begin{minipage}[t]{0.5\linewidth}
  \centering
  \includegraphics[width=3.1in]{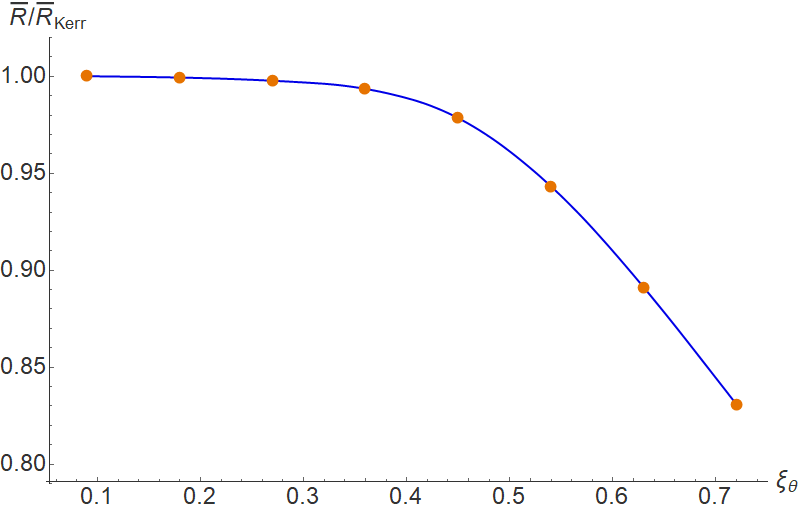}
  \end{minipage}
  }

  \centering
  \caption{The variation of $\bar{R}/\bar{R}_{\text{Kerr}}$ with respect to the opening angle $\xi_\theta$. The spin is fixed at $a=0.998$ and $k_A=16$. Left: $\theta_o=\pi/2$; Right: $\theta_o=17^{\circ}$.}
  \label{comA}
\end{figure}

In Fig. \ref{comA}, we show the variations of $\bar{R}/\bar{R}_{\text{Kerr}}$ with respect to the opening angle $\xi_\theta$ for $\theta_o=\pi/2$ and $\theta_o=17^{\circ}$ respectively. Obviously, we can see that when the observer is located on the equator the value of $\bar{R}/\bar{R}_{\text{Kerr}}$ is much smaller than that when the inclination angle $\theta_o=17^{\circ}$ for the same $\xi_\theta$,  even though $\bar{R}/\bar{R}_{\text{Kerr}}$ changes  little with the increase of $\xi_\theta$. The main reason is that the distribution density of the plasma along the polar direction is the largest at $\theta_o=\pi/2$,  and the photons must travel through the thickest plasma. Meanwhile, roughly speaking the size of shadow can be regarded as the complement of photon-escaping cone centred at the observer. Thus, the region of the plasma whose polar angle is near the value of the inclination angle $\theta_o$ is crucial to the formation of a black hole shadow in the sight of the observers, and the plasma with a higher density would play a greater influence on the trajectories of photons.

\begin{figure}[h!]
  \centering

  \subfigure[$k_A=1$]{
  \begin{minipage}[t]{0.3\linewidth}
  \centering
  \includegraphics[width=1.5in]{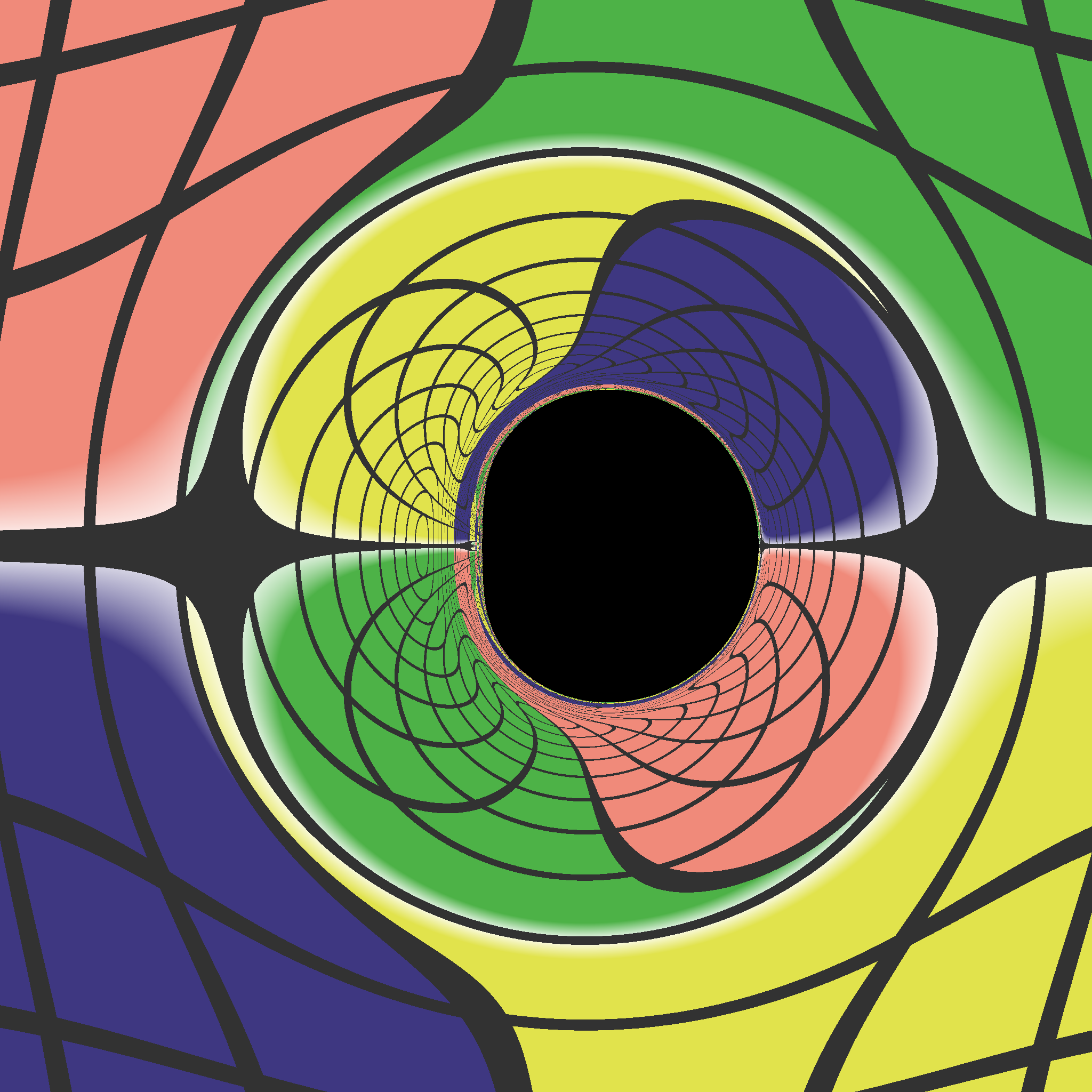}
  \end{minipage}%
  }%
\subfigure[$k_A=11$]{
  \begin{minipage}[t]{0.3\linewidth}
  \centering
  \includegraphics[width=1.5in]{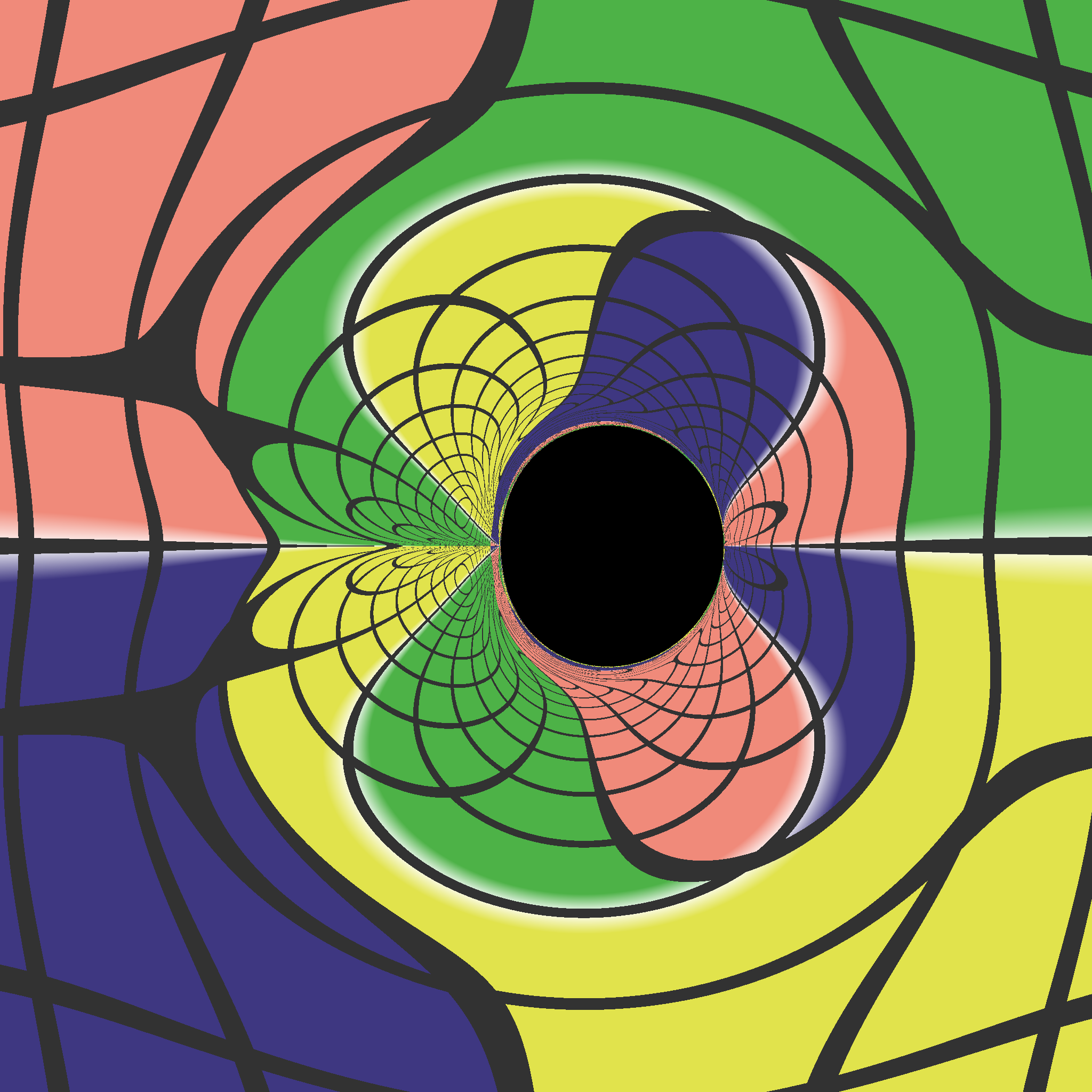}
  \end{minipage}
  }%
\subfigure[$k_A=26$]{
  \begin{minipage}[t]{0.3\linewidth}
  \centering
  \includegraphics[width=1.5in]{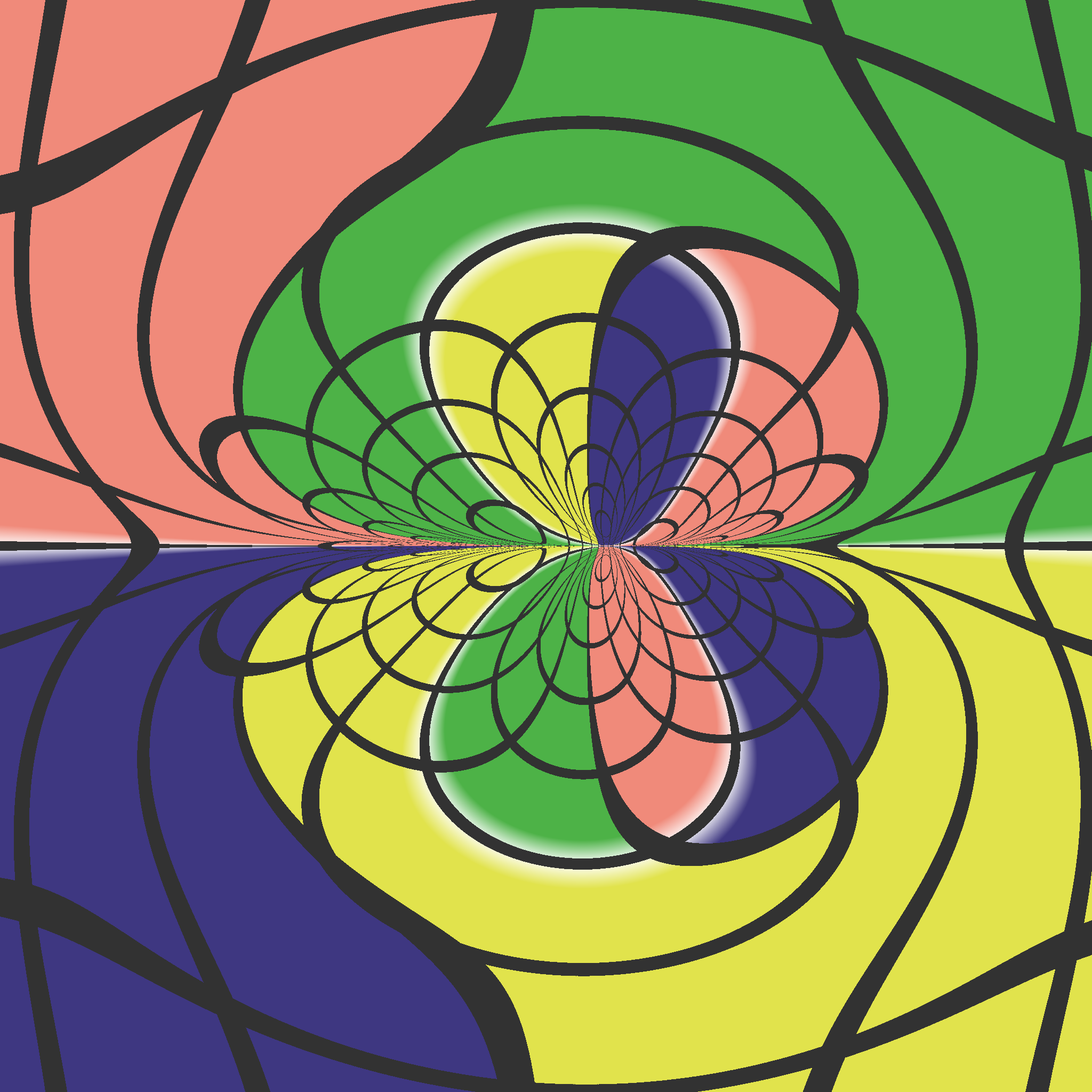}
  \end{minipage}
  }%
  
  \centering
  \caption{The images of the Kerr black hole surrounded by a plasma of model A with different $k_A$. The inclination angle of the observer is fixed at $\theta_o=\pi/2$. The spin is fixed at $a=0.998$ and $\xi_\theta=0.36$.}
  \label{mAka}
\end{figure}

Next, we fix $\xi_\theta=0.36$ and vary $k_A$ to see the influence of $k_A$ on the black hole shadow. The results can be found in Fig. \ref{mAka}, where we place the observer on the equatorial plane. One can see the size of the black hole shadow becomes smaller as $k_A$ goes up, and even the black hole shadow would disappear when $k_A$ is big enough. Our results are consistent with those in \cite{Huang:2018rfn}. 

\subsection{Model B}

\begin{figure}[p!]
  \centering

  \subfigure[$i=2, r_1\simeq0.2$]{
  \begin{minipage}[t]{0.19\linewidth}
  \centering
  \includegraphics[width=1.2in]{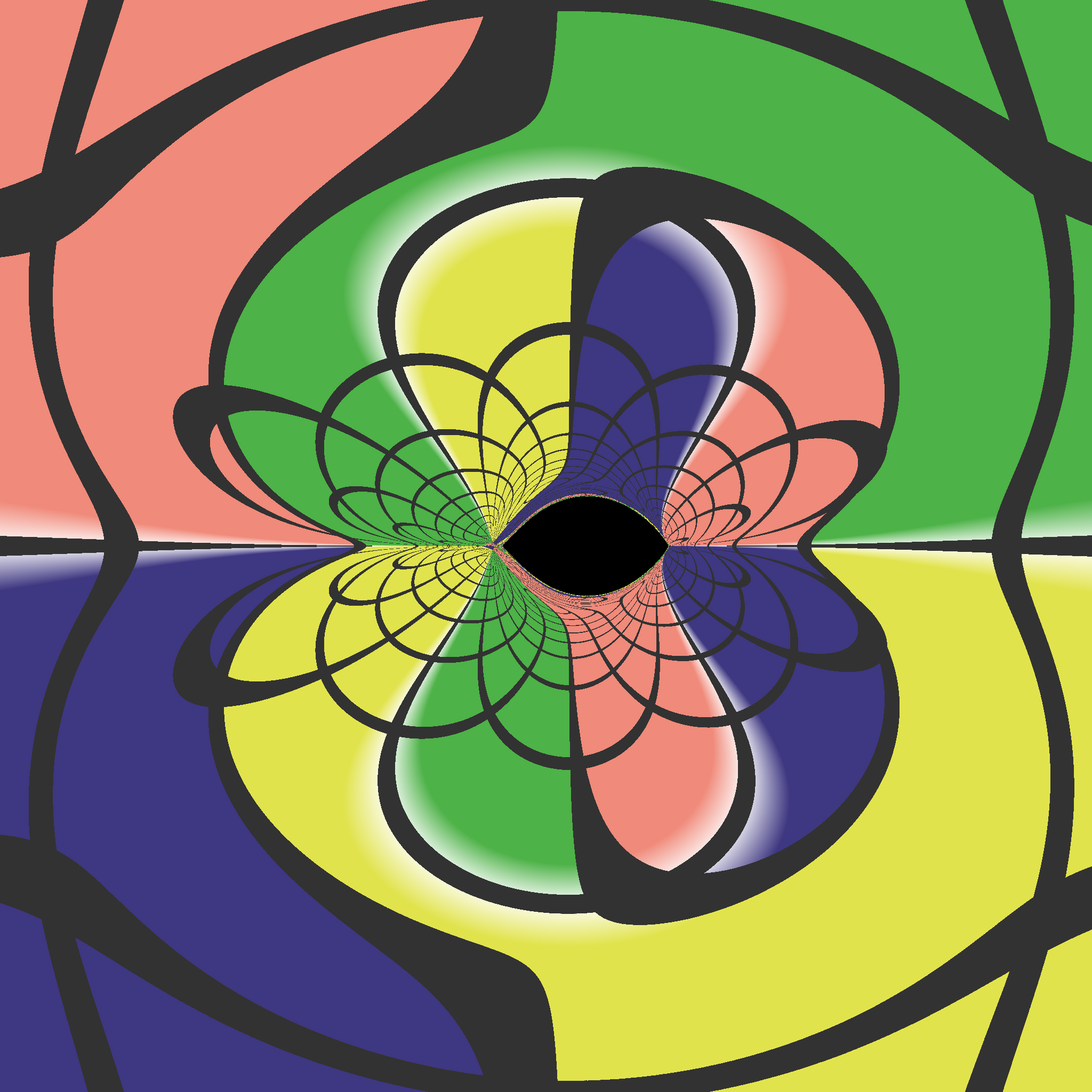}
  \end{minipage}%
  }%
  \subfigure[$i=4, r_1\simeq0.4$]{
  \begin{minipage}[t]{0.19\linewidth}
  \centering
  \includegraphics[width=1.2in]{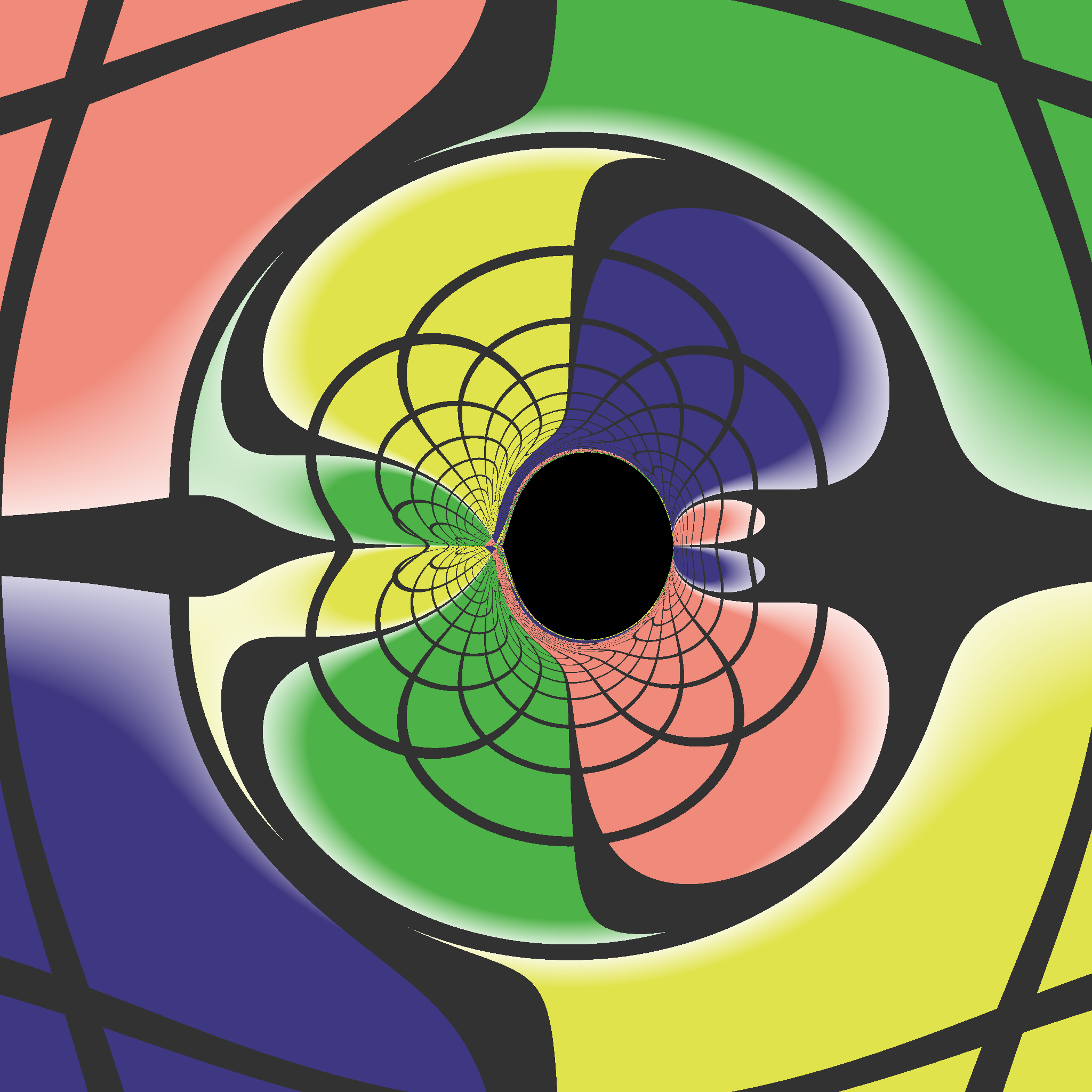}
  \end{minipage}%
  }%
  \subfigure[$i=6, r_1\simeq0.6$]{
  \begin{minipage}[t]{0.19\linewidth}
  \centering
  \includegraphics[width=1.2in]{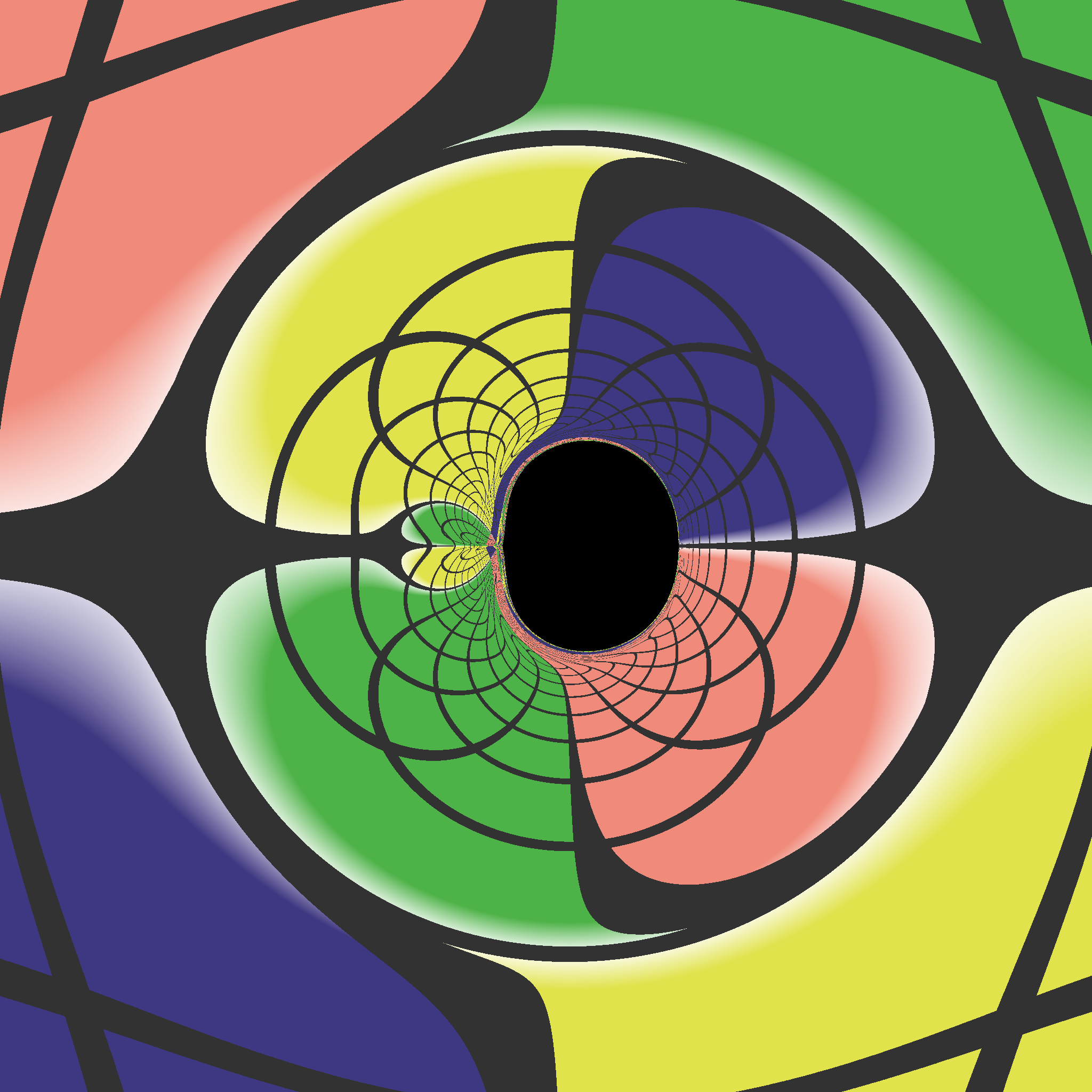}
  \end{minipage}
  }%
  \subfigure[$i=8, r_1\simeq0.8$]{
  \begin{minipage}[t]{0.19\linewidth}
  \centering
  \includegraphics[width=1.2in]{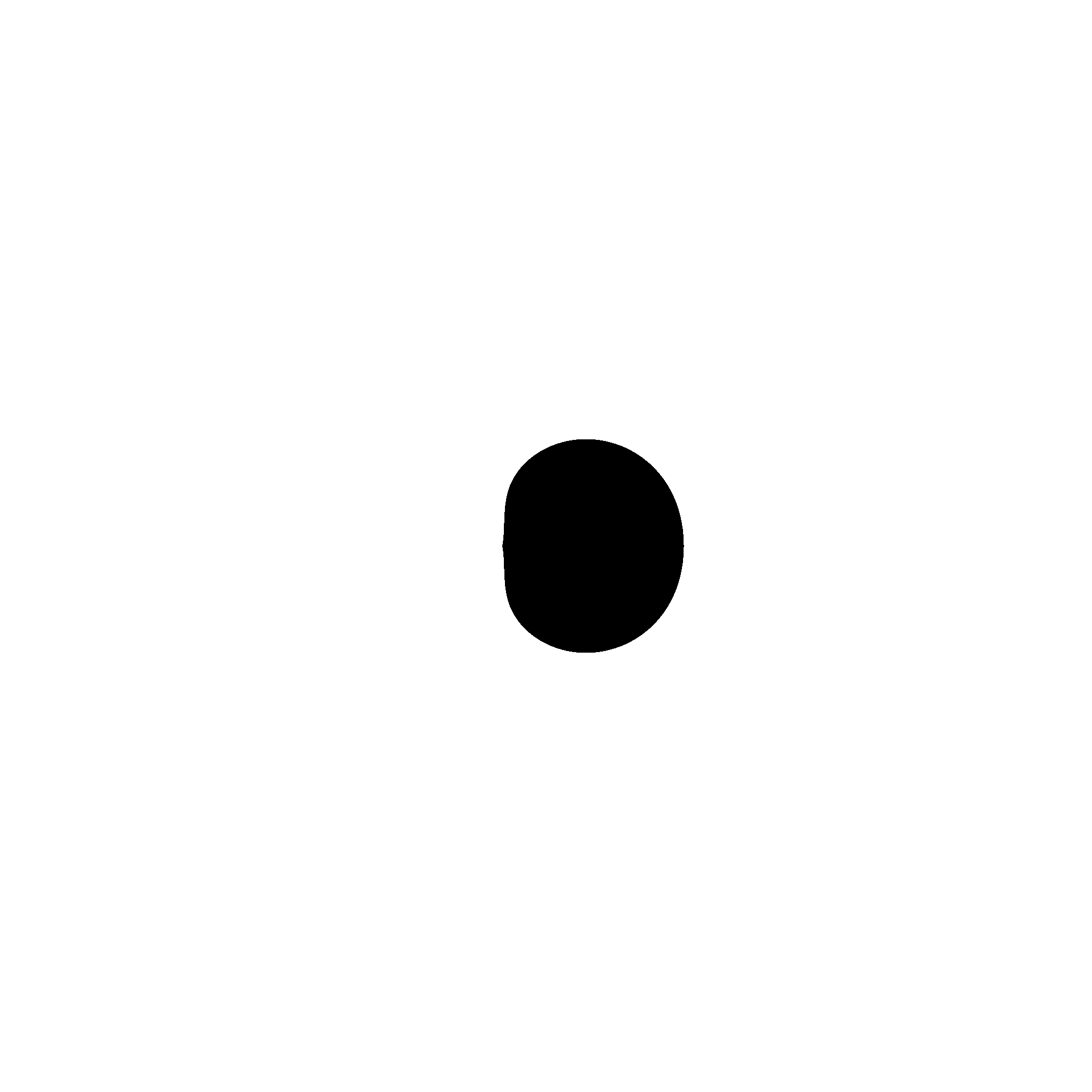}
  \end{minipage}
  }%
  \subfigure[$i=10, r_1\simeq1.0$]{
  \begin{minipage}[t]{0.19\linewidth}
  \centering
  \includegraphics[width=1.2in]{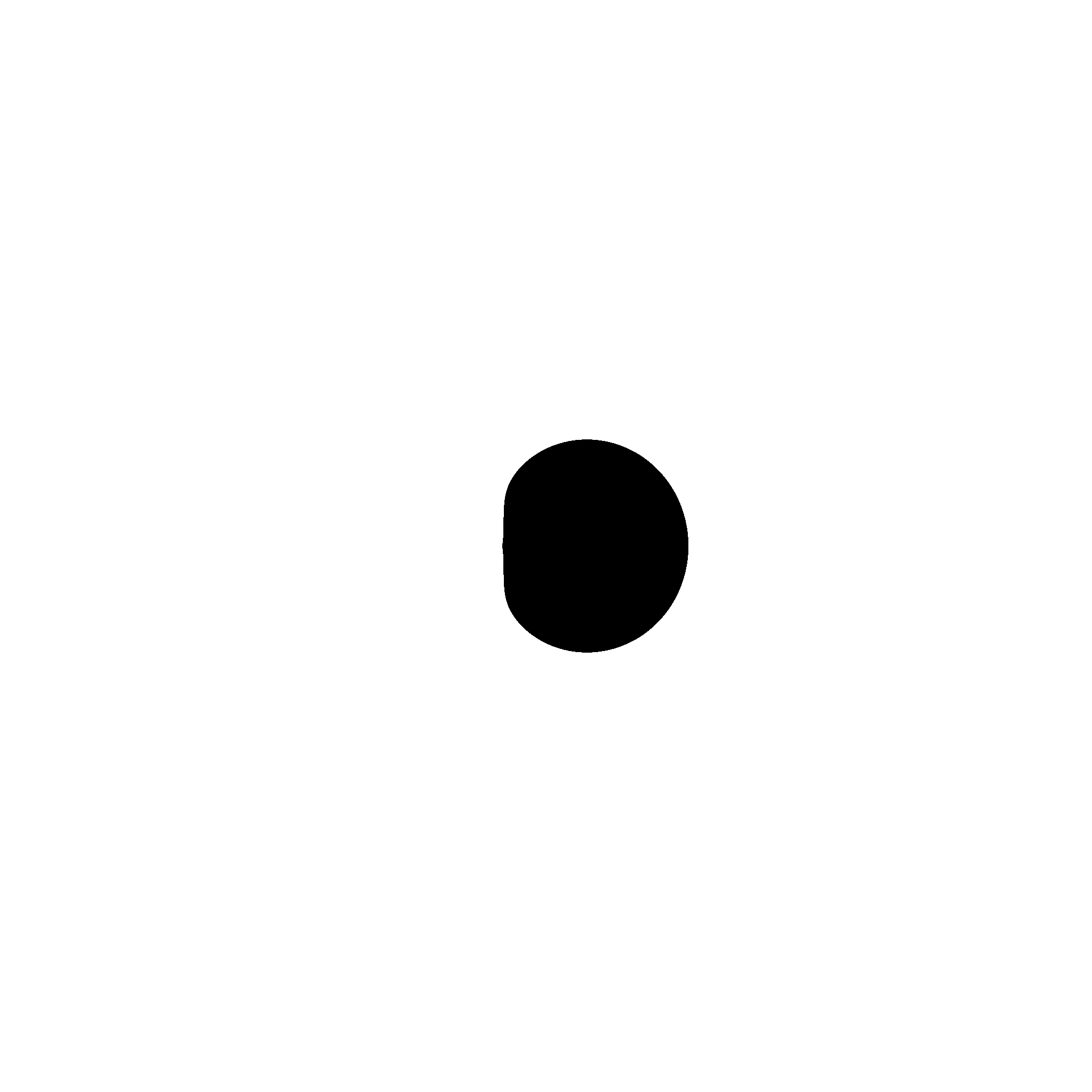}
  \end{minipage}
  }%
  
     \subfigure[$i=2, r_1\simeq0.6$]{
  \begin{minipage}[t]{0.19\linewidth}
  \centering
  \includegraphics[width=1.2in]{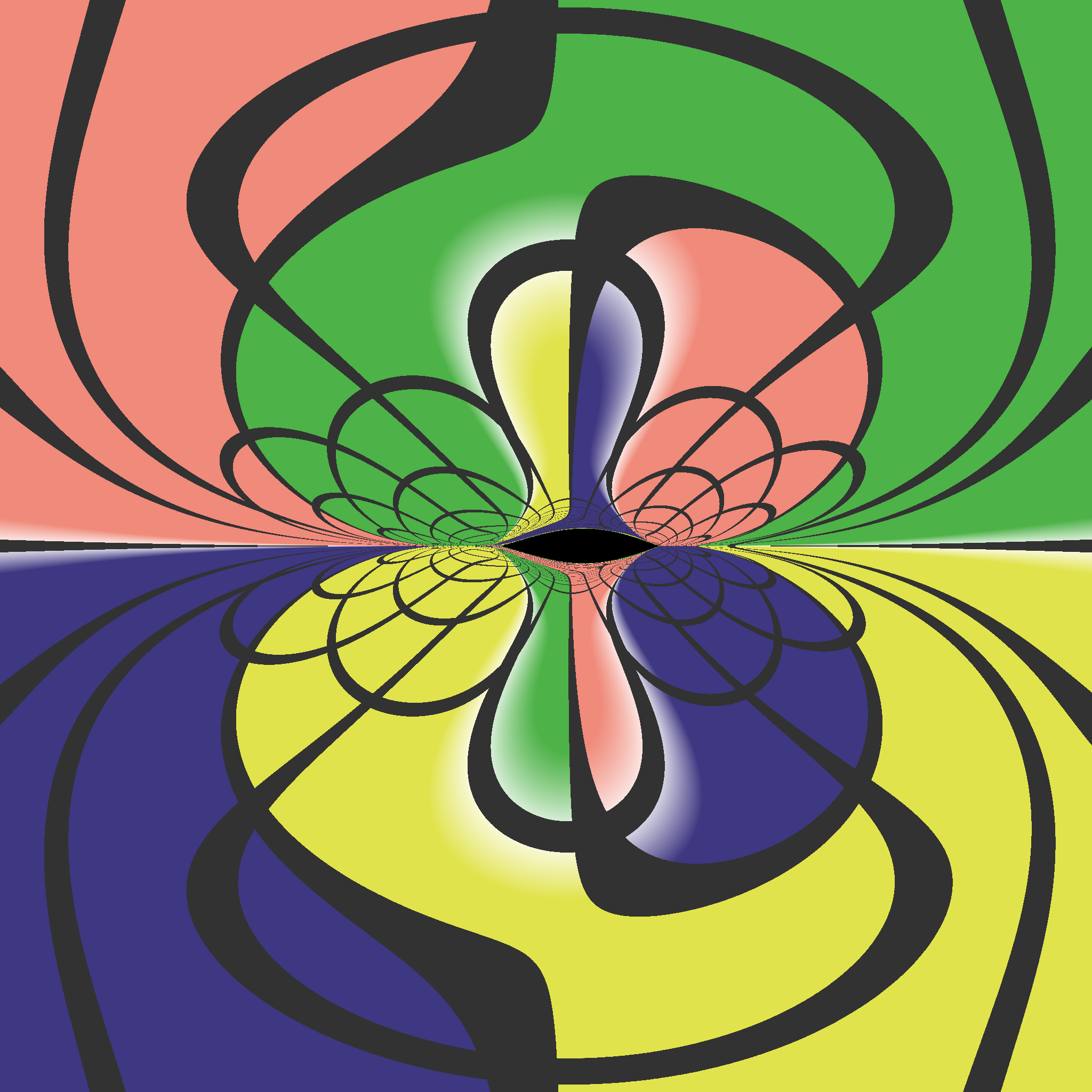}
  \end{minipage}%
  }%
  \subfigure[$i=4, r_1\simeq1.2$]{
  \begin{minipage}[t]{0.19\linewidth}
  \centering
  \includegraphics[width=1.2in]{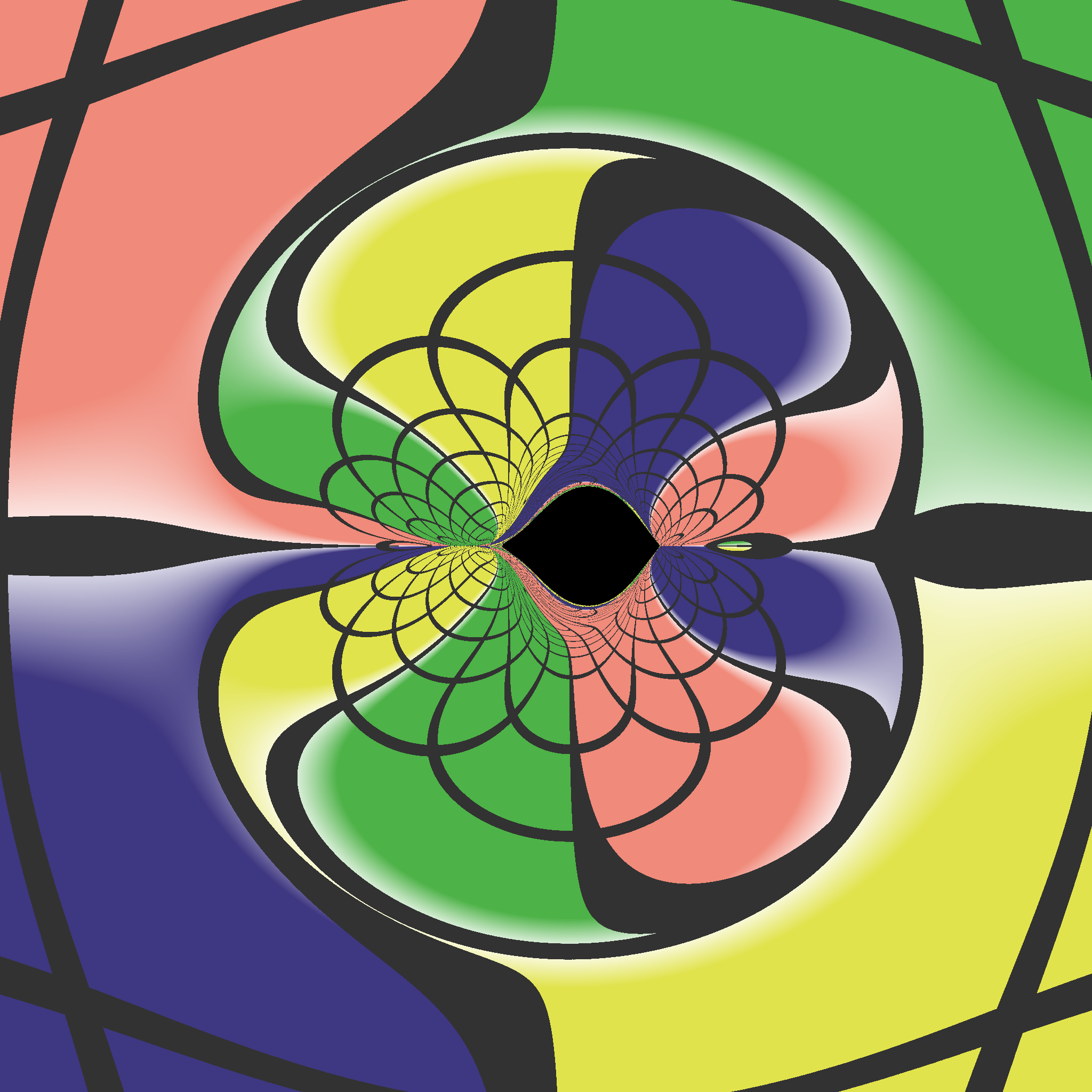}
  \end{minipage}%
  }%
  \subfigure[$i=6, r_1\simeq1.8$]{
  \begin{minipage}[t]{0.19\linewidth}
  \centering
  \includegraphics[width=1.2in]{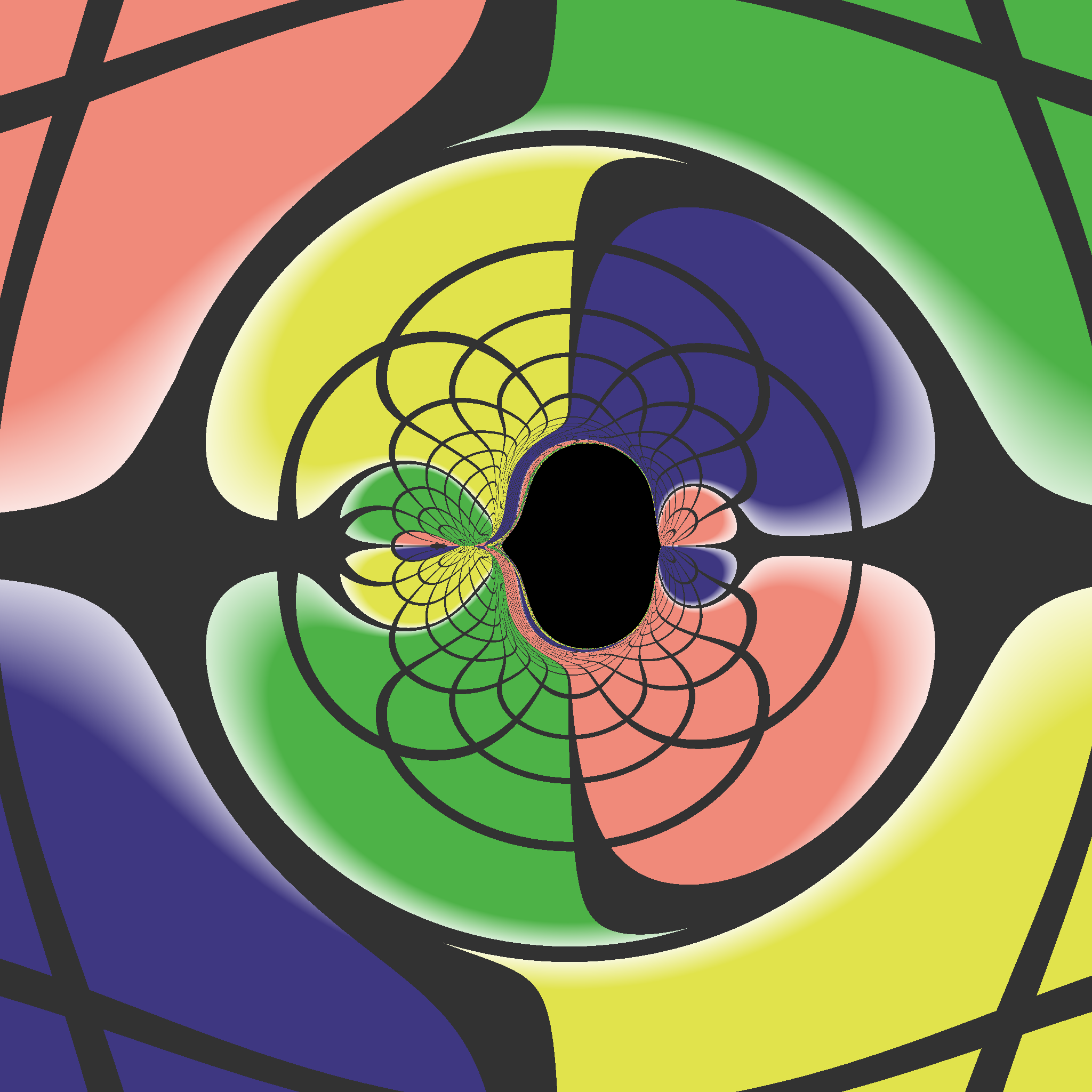}
  \end{minipage}
  }%
  \subfigure[$i=8, r_1\simeq2.4$]{
  \begin{minipage}[t]{0.19\linewidth}
  \centering
  \includegraphics[width=1.2in]{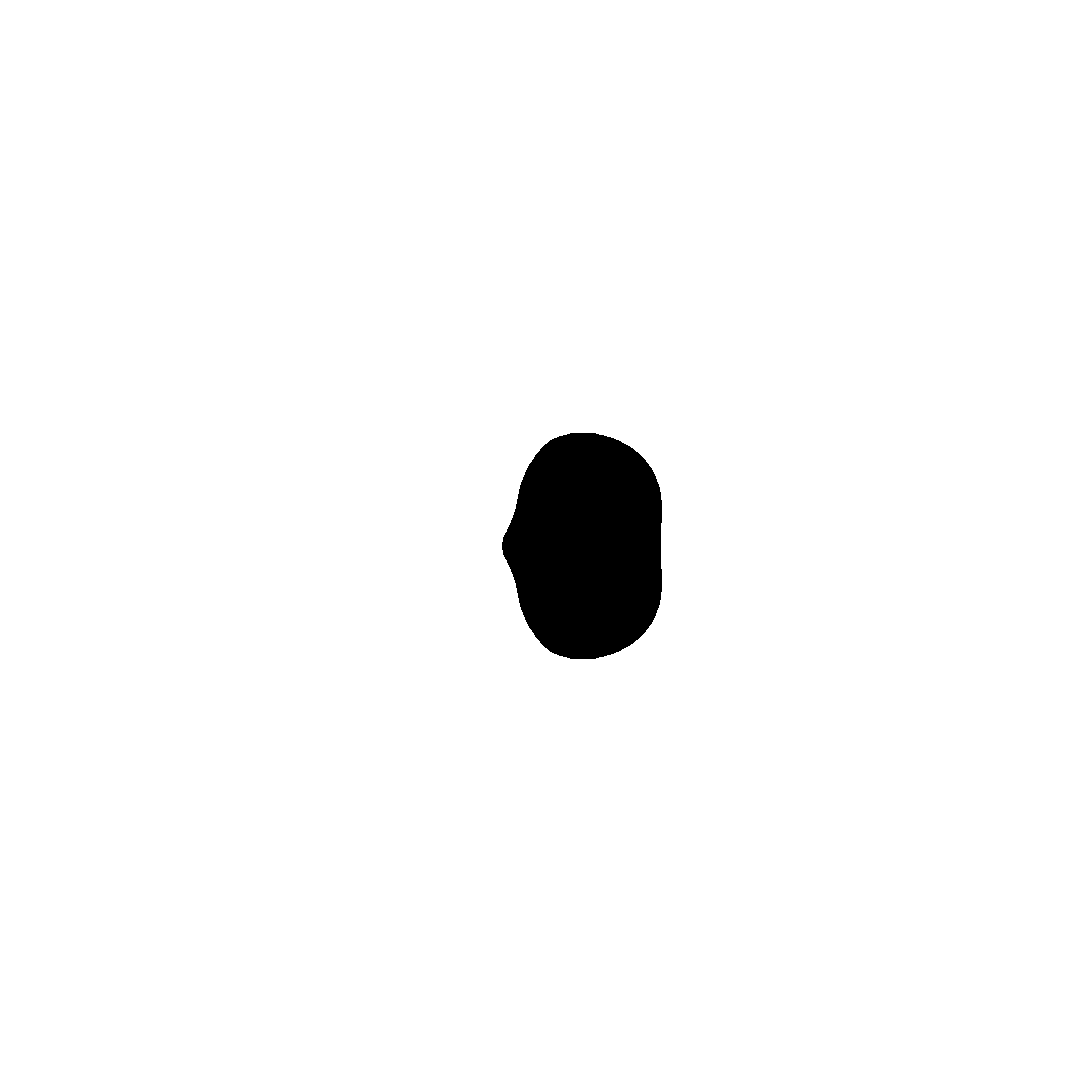}
  \end{minipage}
  }%
  \subfigure[$i=10, r_1\simeq3.0$]{
  \begin{minipage}[t]{0.19\linewidth}
  \centering
  \includegraphics[width=1.2in]{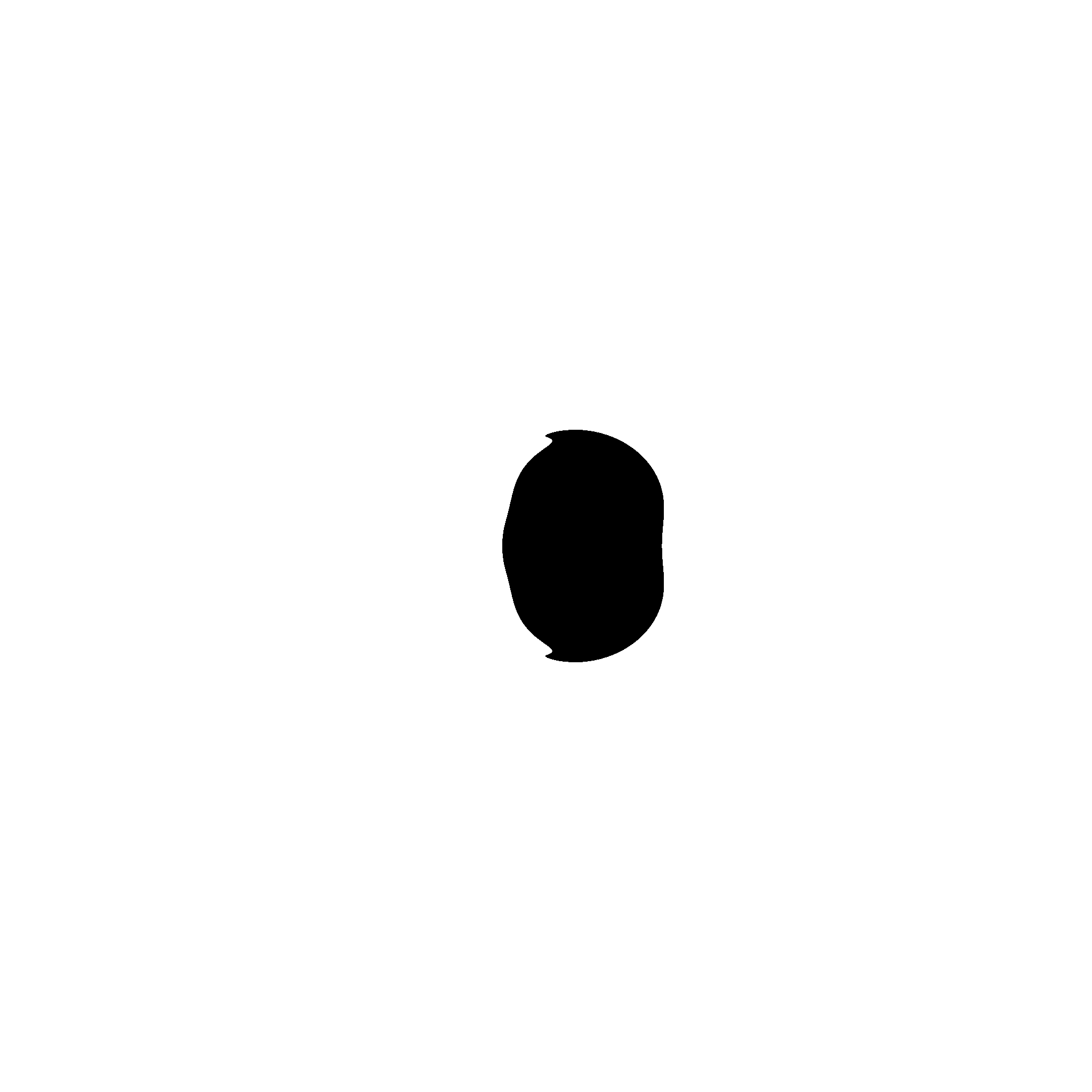}
  \end{minipage}
  }%
  
    \subfigure[$i=2, r_1\simeq1.0$]{
  \begin{minipage}[t]{0.19\linewidth}
  \centering
  \includegraphics[width=1.2in]{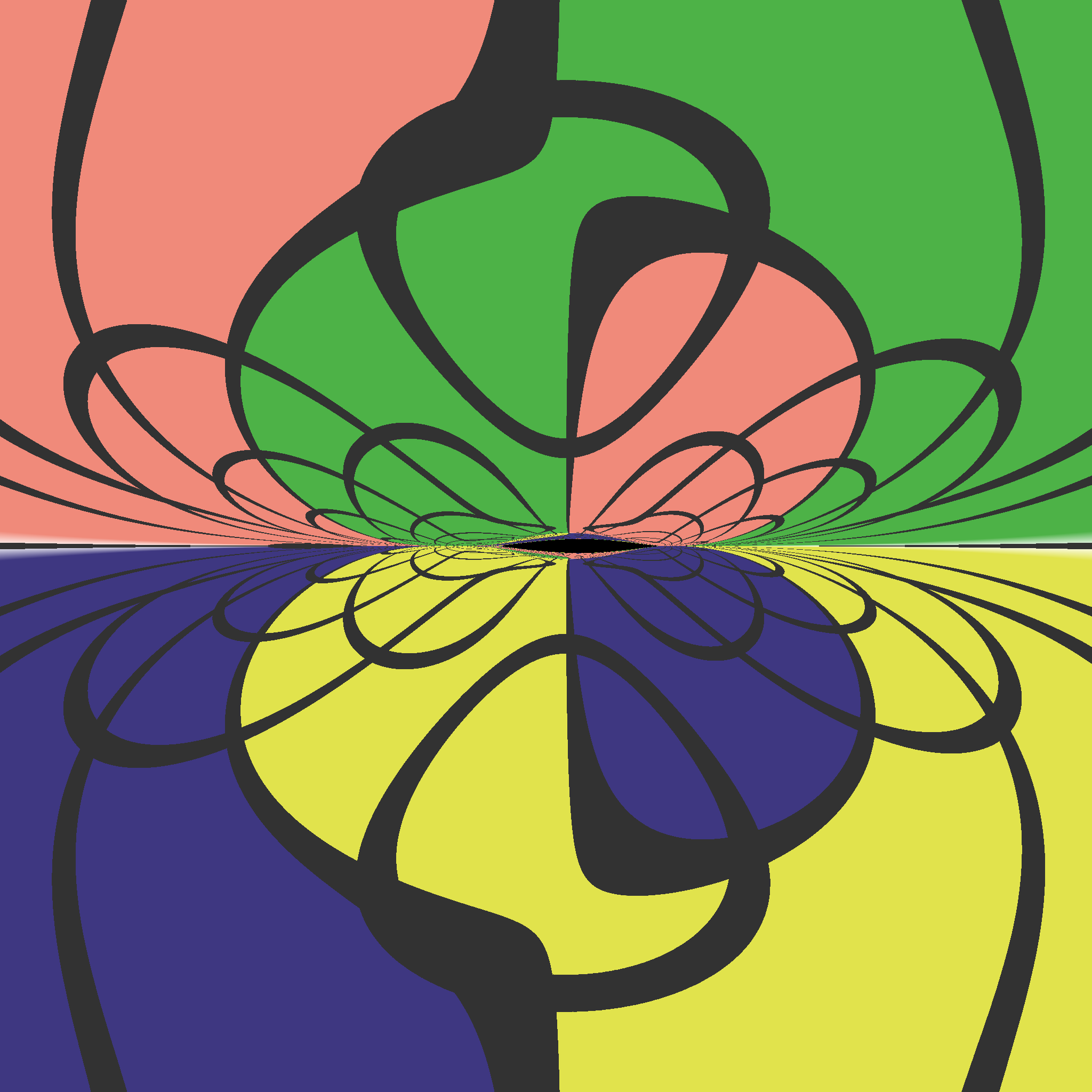}
  \end{minipage}%
  }%
  \subfigure[$i=4, r_1\simeq2.0$]{
  \begin{minipage}[t]{0.19\linewidth}
  \centering
  \includegraphics[width=1.2in]{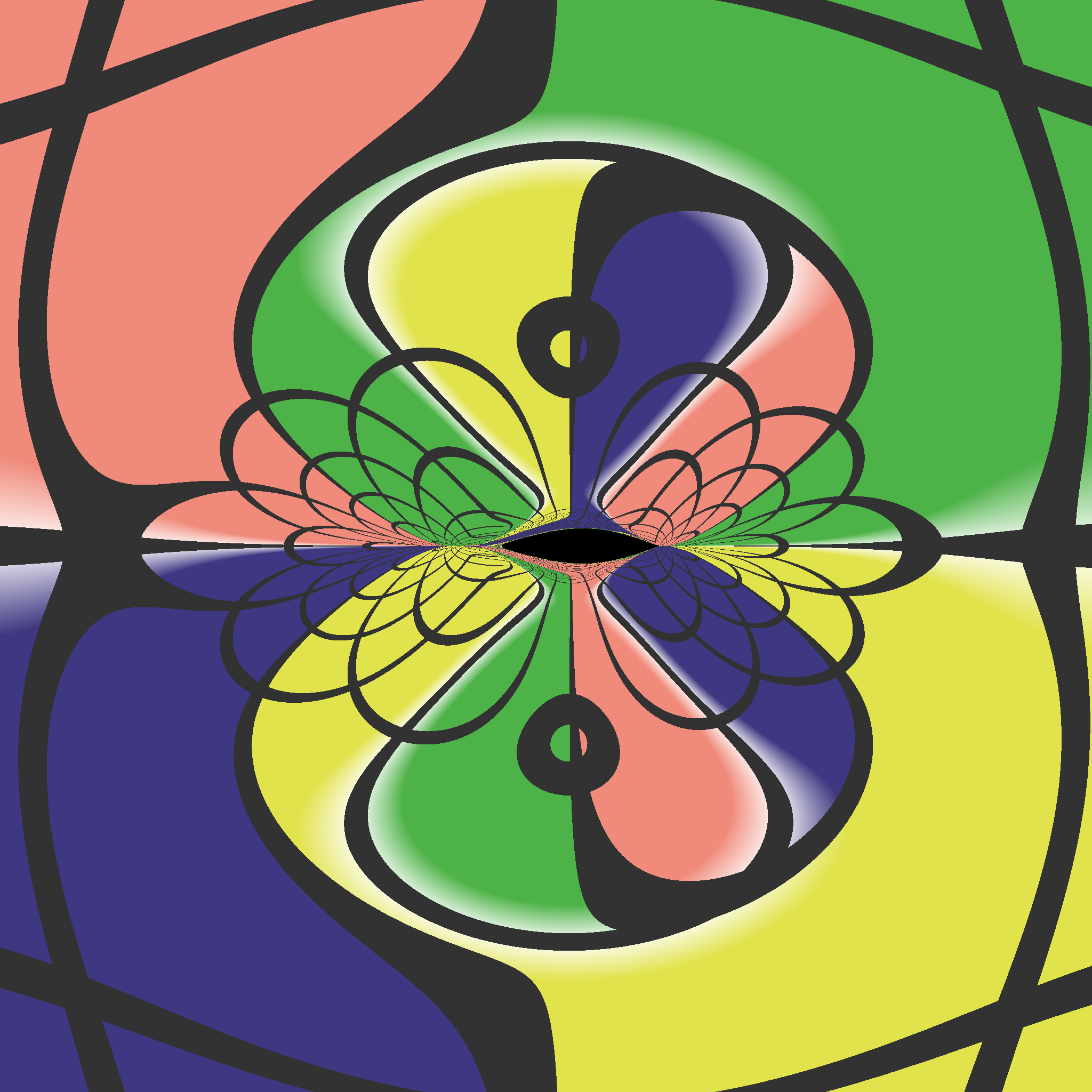}
  \end{minipage}%
  }%
  \subfigure[$i=6, r_1\simeq3.0$]{
  \begin{minipage}[t]{0.19\linewidth}
  \centering
  \includegraphics[width=1.2in]{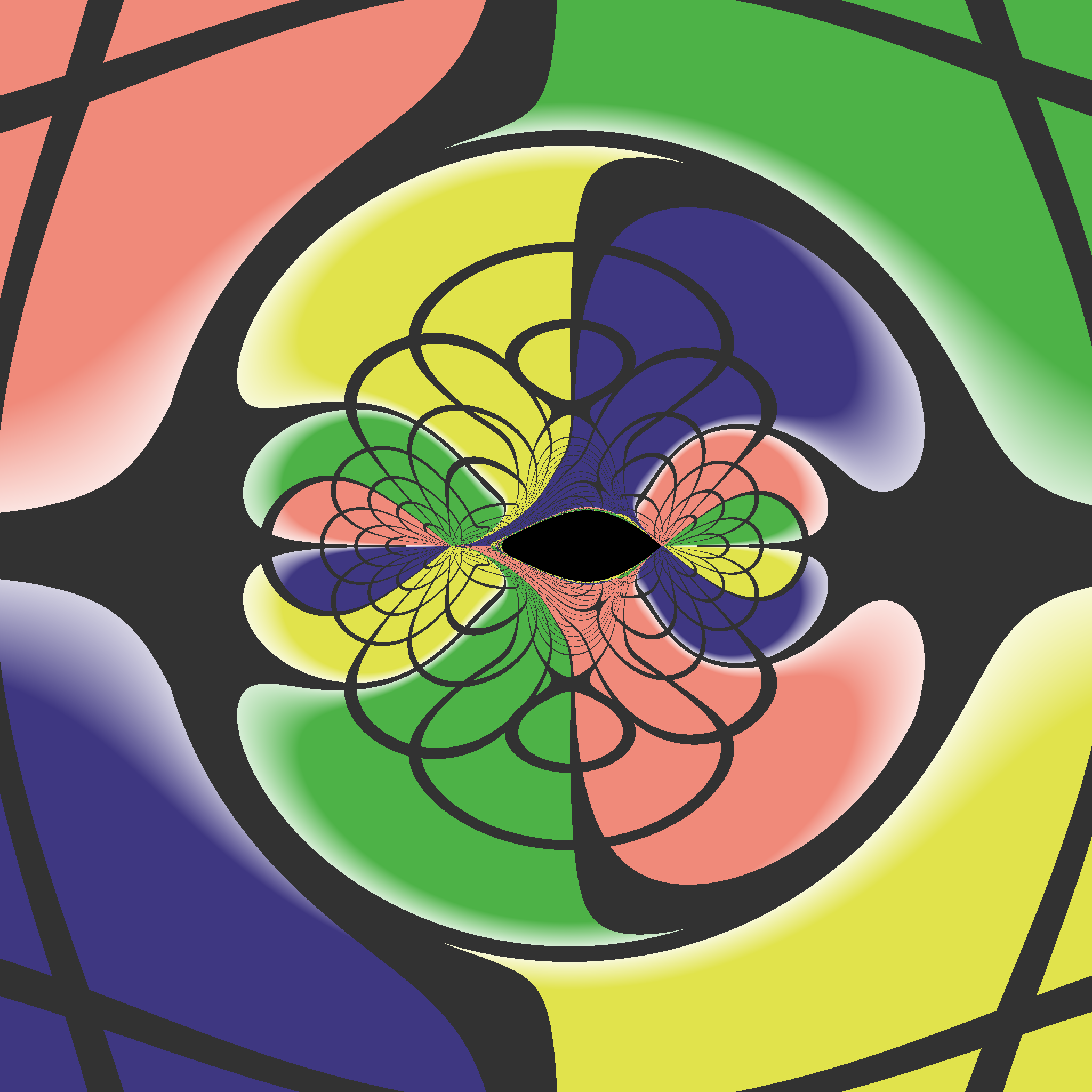}
  \end{minipage}
  }%
  \subfigure[$i=8, r_1\simeq4.0$]{
  \begin{minipage}[t]{0.19\linewidth}
  \centering
  \includegraphics[width=1.2in]{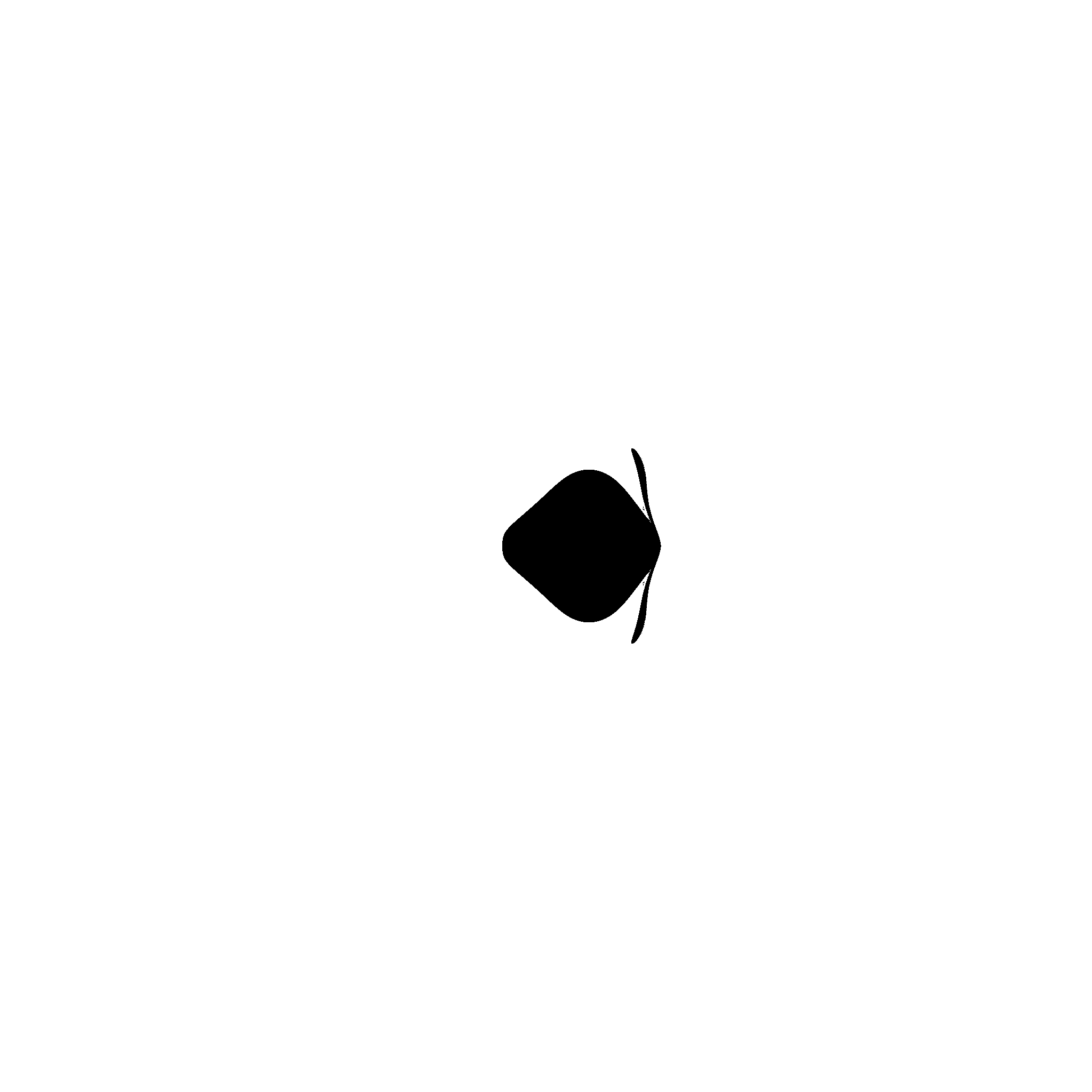}
  \end{minipage}
  }%
  \subfigure[$i=10, r_1\simeq5.0$]{
  \begin{minipage}[t]{0.19\linewidth}
  \centering
  \includegraphics[width=1.2in]{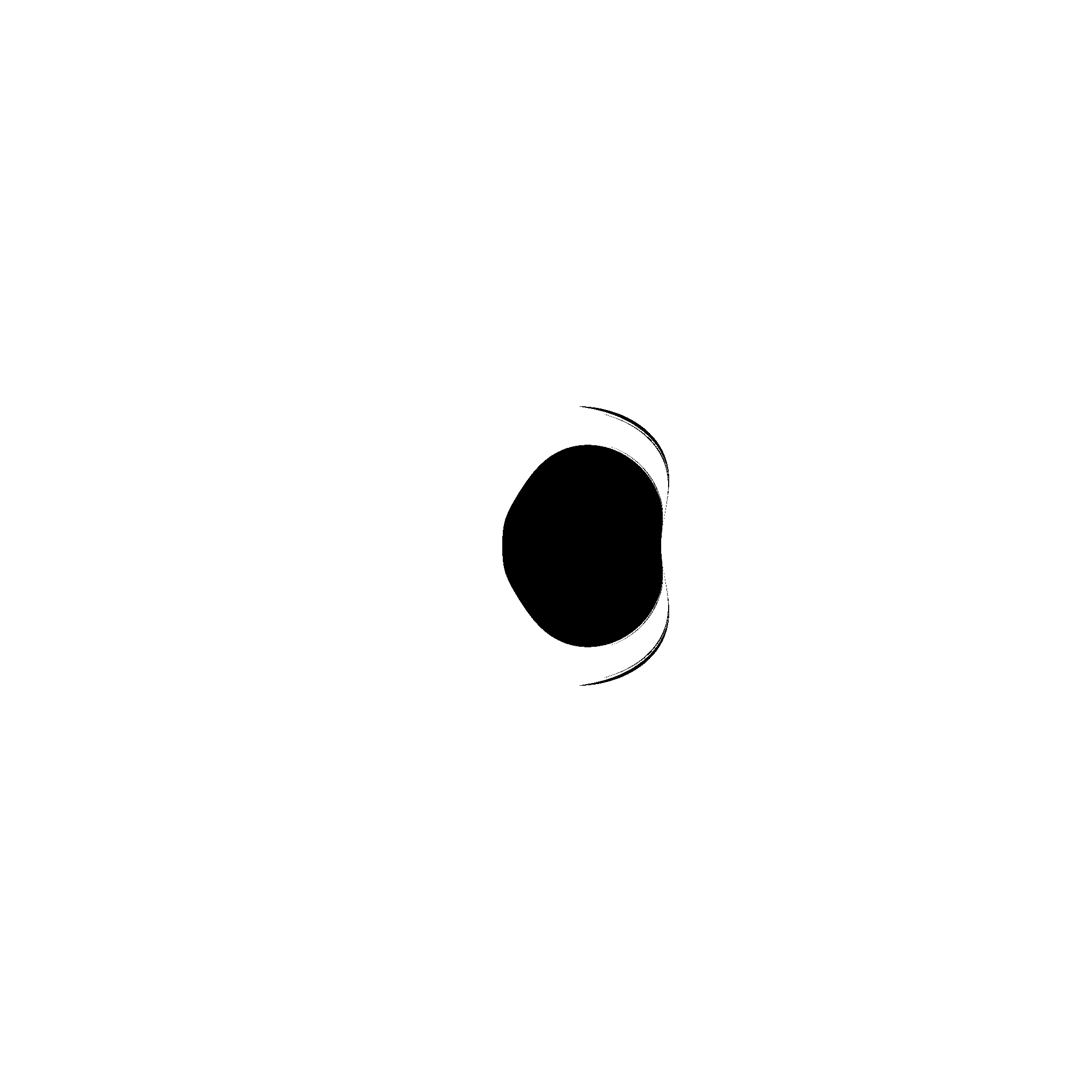}
  \end{minipage}
  }%

    \subfigure[$i=2, r_1\simeq1.4$]{
  \begin{minipage}[t]{0.19\linewidth}
  \centering
  \includegraphics[width=1.2in]{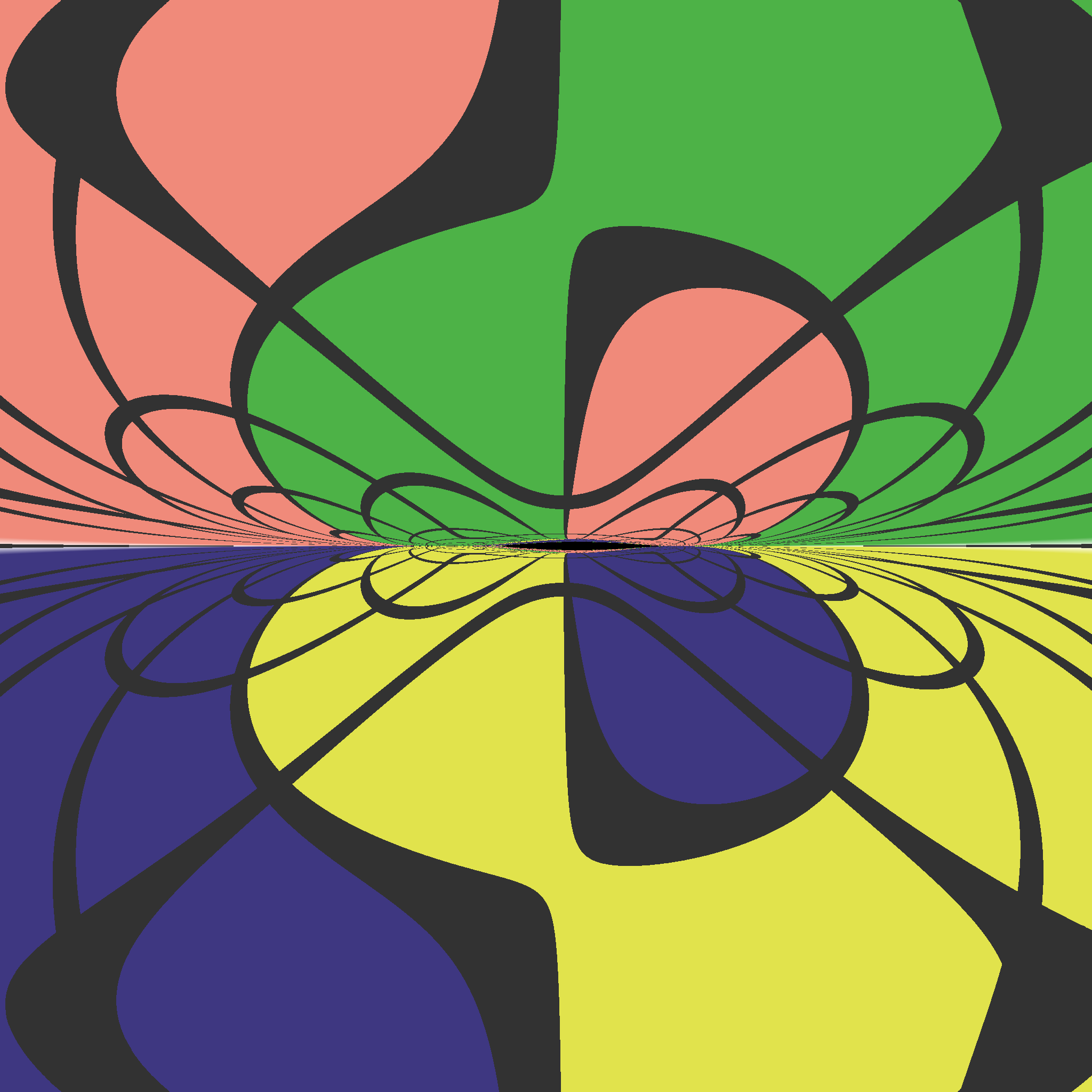}
  \end{minipage}%
  }%
  \subfigure[$i=4, r_1\simeq2.8$]{
  \begin{minipage}[t]{0.19\linewidth}
  \centering
  \includegraphics[width=1.2in]{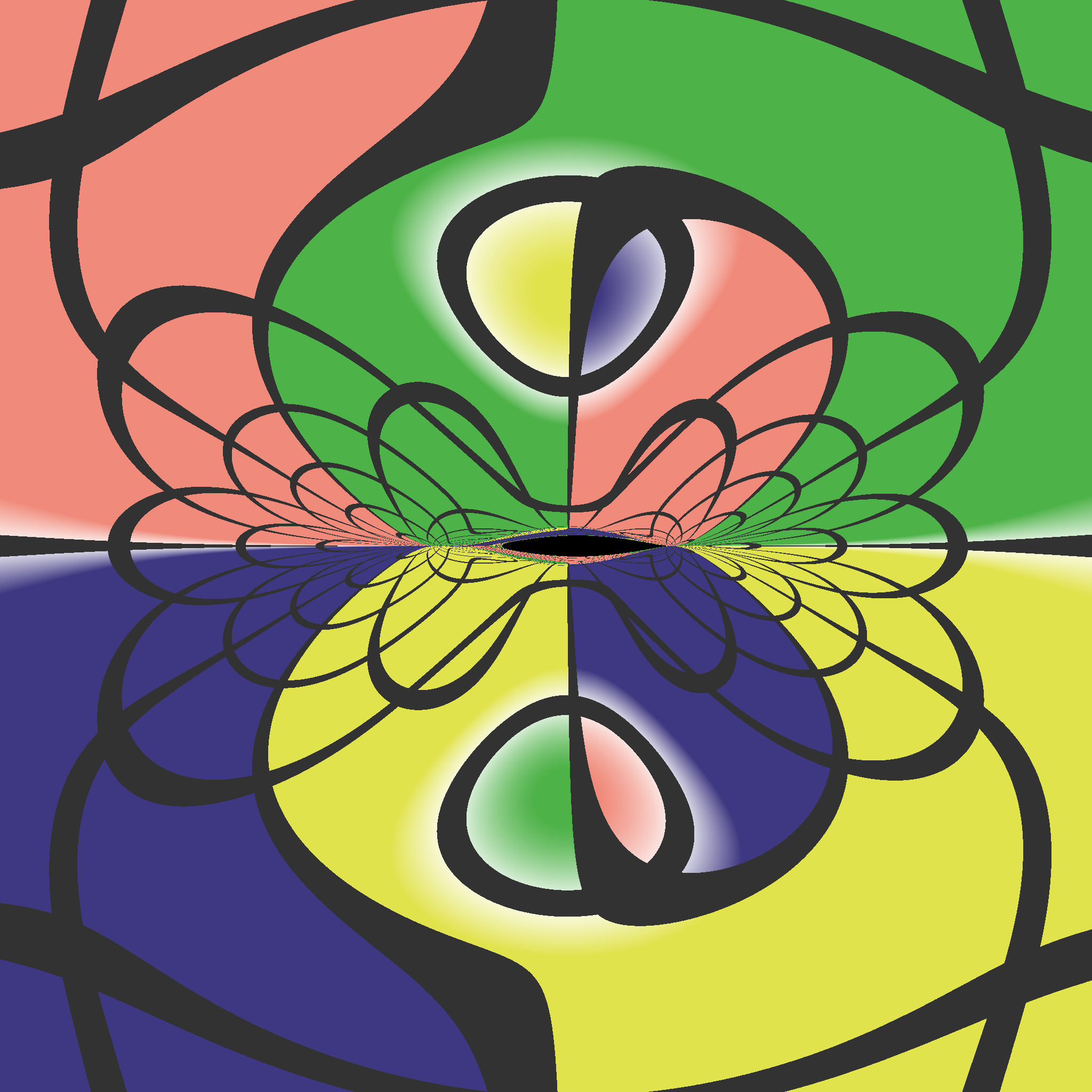}
  \end{minipage}%
  }%
  \subfigure[$i=6, r_1\simeq4.2$]{
  \begin{minipage}[t]{0.19\linewidth}
  \centering
  \includegraphics[width=1.2in]{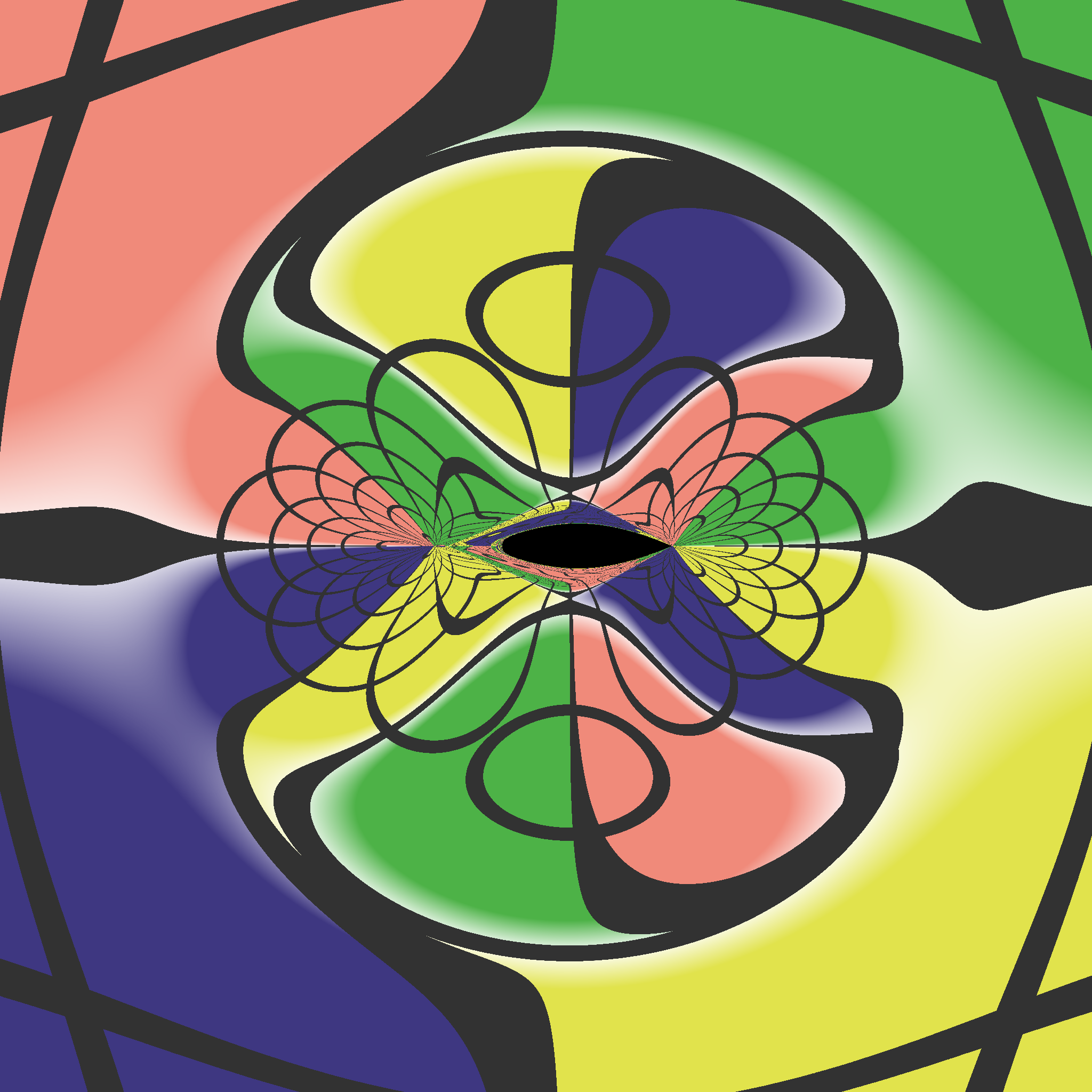}
  \end{minipage}
  }%
  \subfigure[$i=8, r_1\simeq5.6$]{
  \begin{minipage}[t]{0.19\linewidth}
  \centering
  \includegraphics[width=1.2in]{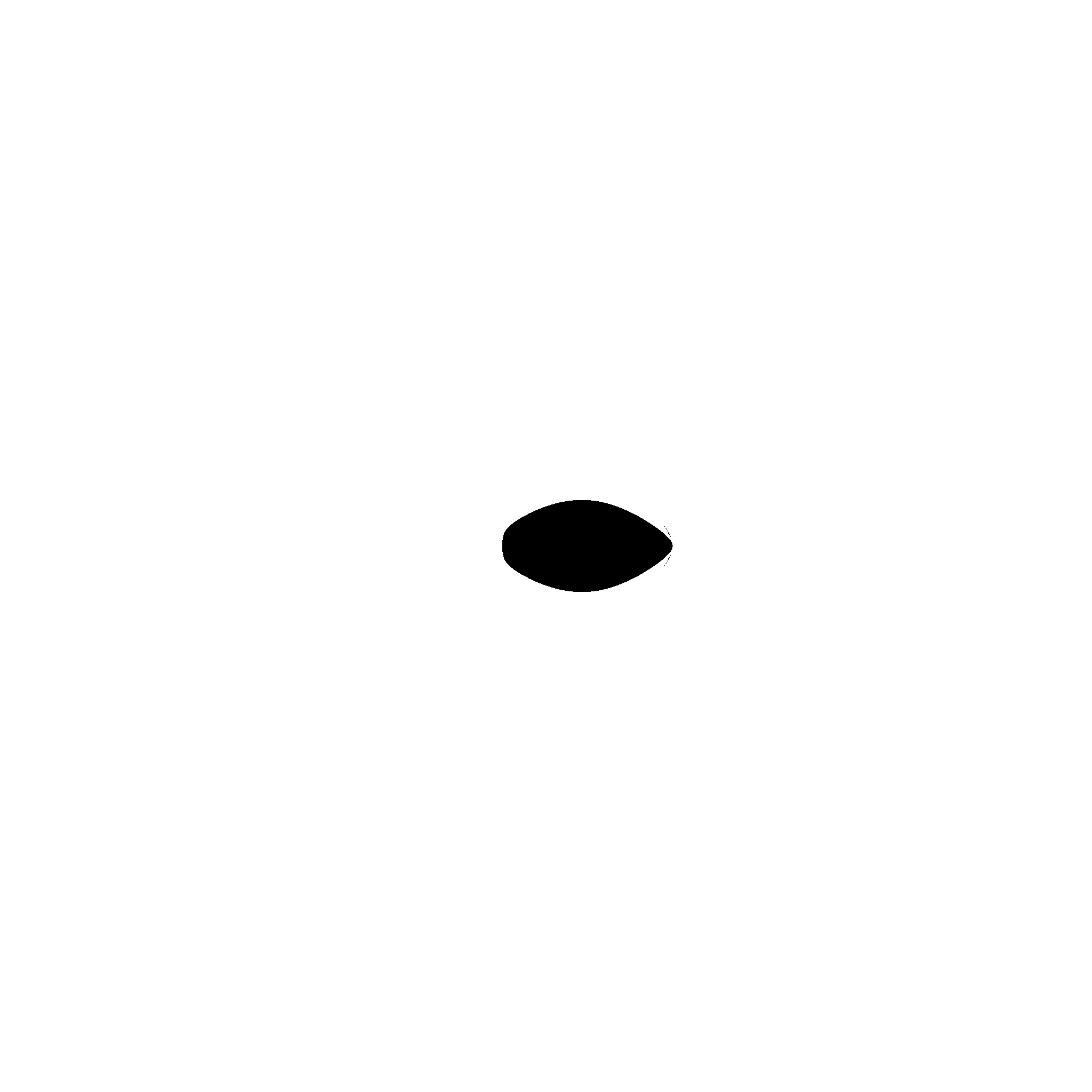}
  \end{minipage}
  }%
  \subfigure[$i=10, r_1\simeq7.0$]{
  \begin{minipage}[t]{0.19\linewidth}
  \centering
  \includegraphics[width=1.2in]{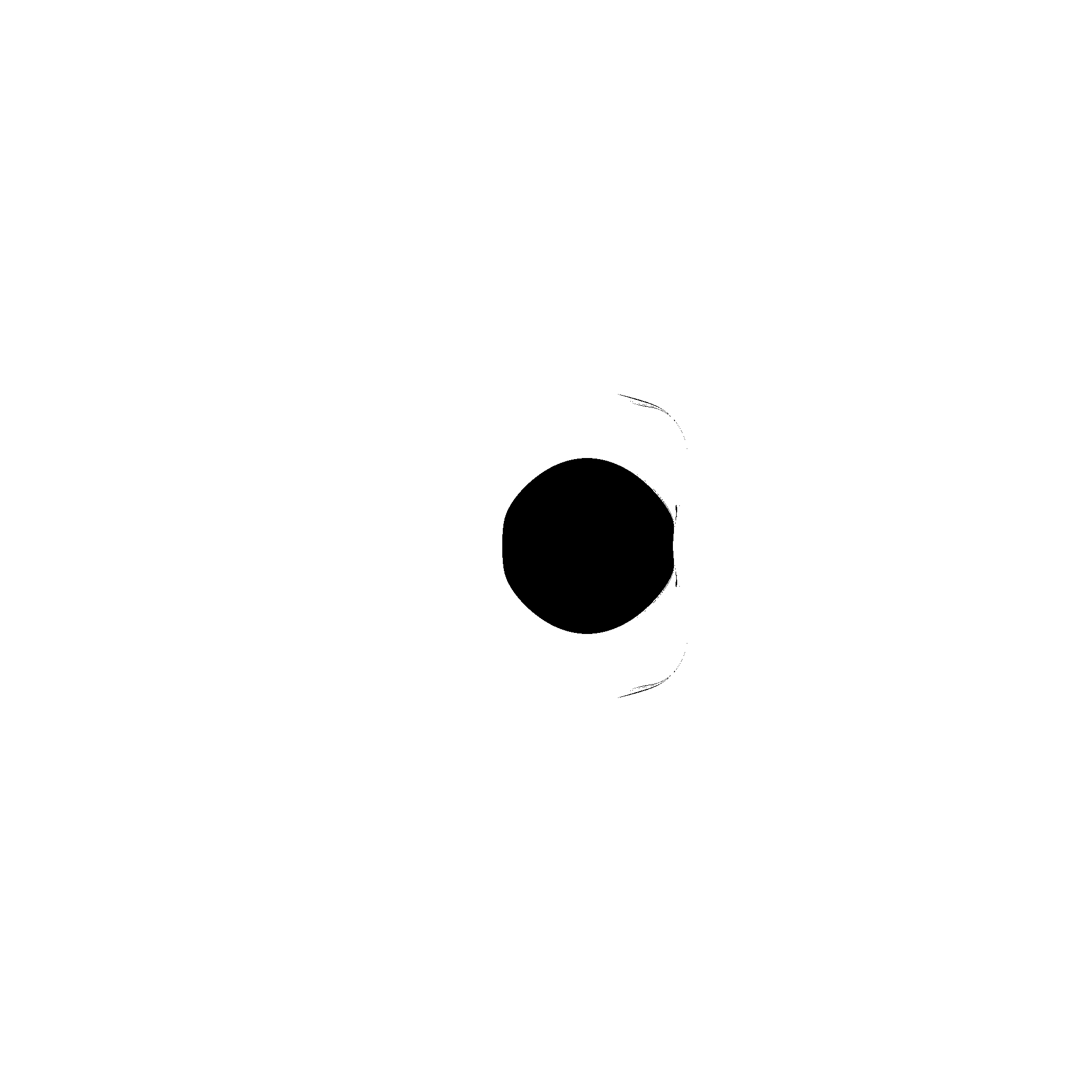}
  \end{minipage}
  }%
  
    \subfigure[$i=2, r_1\simeq1.8$]{
  \begin{minipage}[t]{0.19\linewidth}
  \centering
  \includegraphics[width=1.2in]{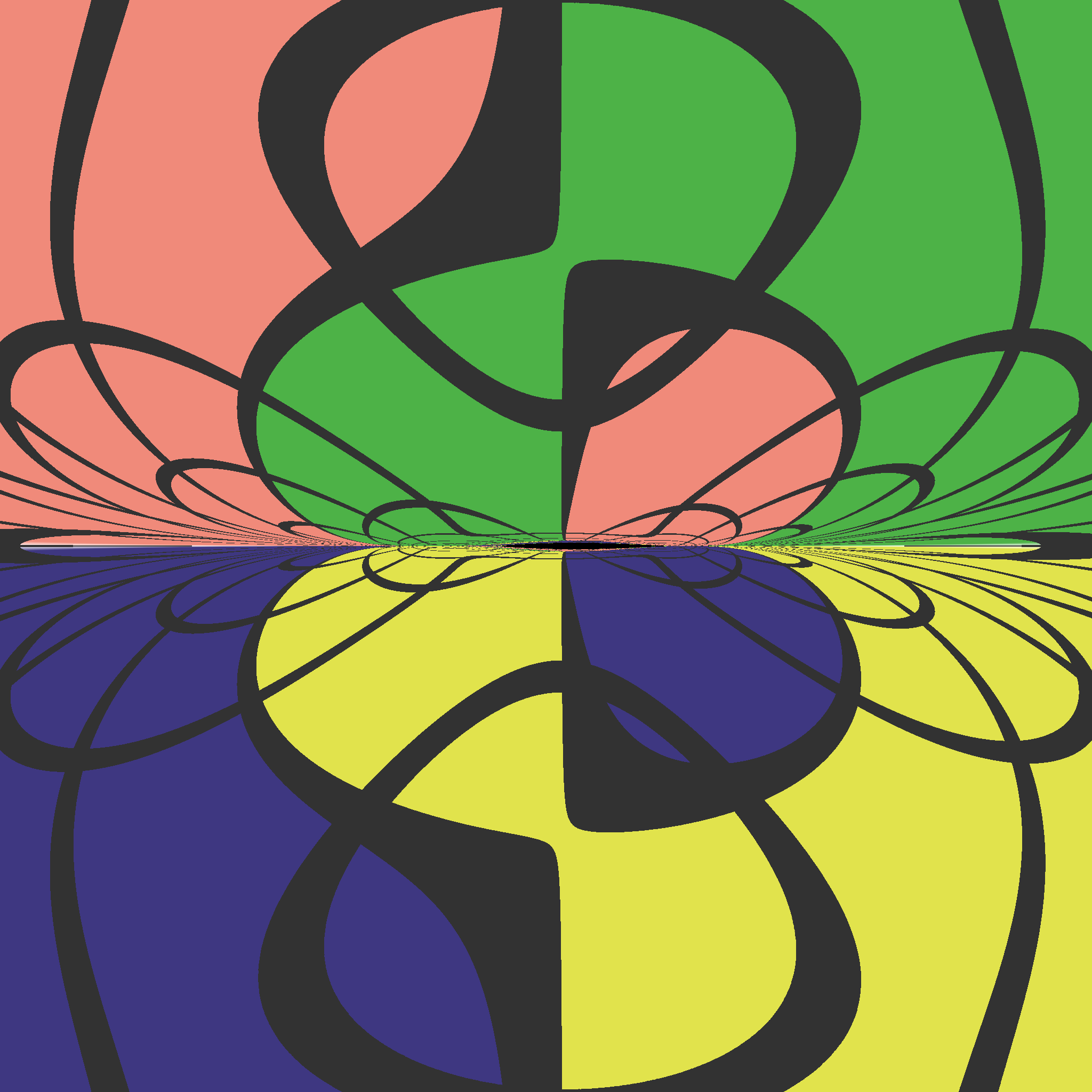}
  \end{minipage}%
  }%
  \subfigure[$i=4, r_1\simeq3.6$]{
  \begin{minipage}[t]{0.19\linewidth}
  \centering
  \includegraphics[width=1.2in]{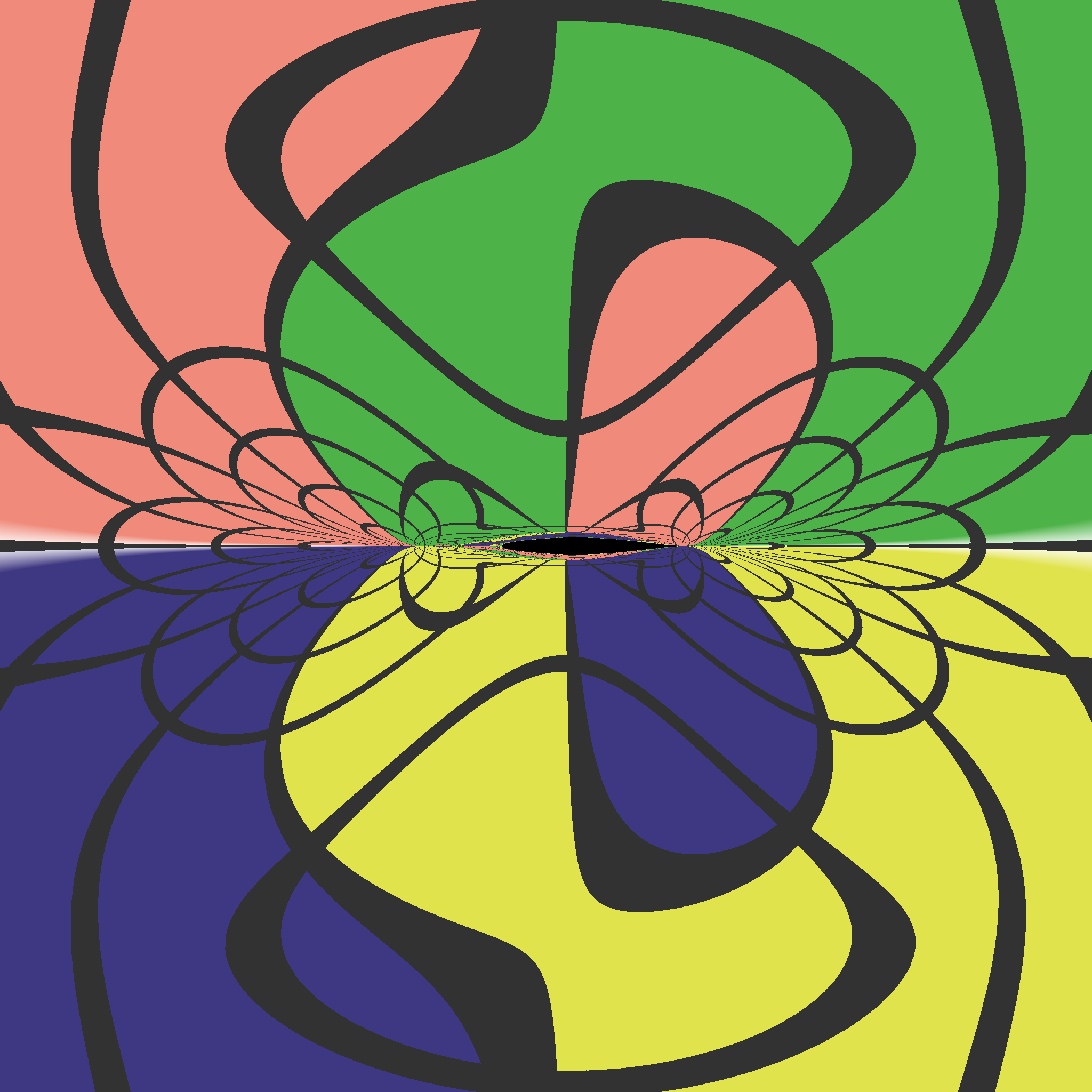}
  \end{minipage}%
  }%
  \subfigure[$i=6, r_1\simeq5.4$]{
  \begin{minipage}[t]{0.19\linewidth}
  \centering
  \includegraphics[width=1.2in]{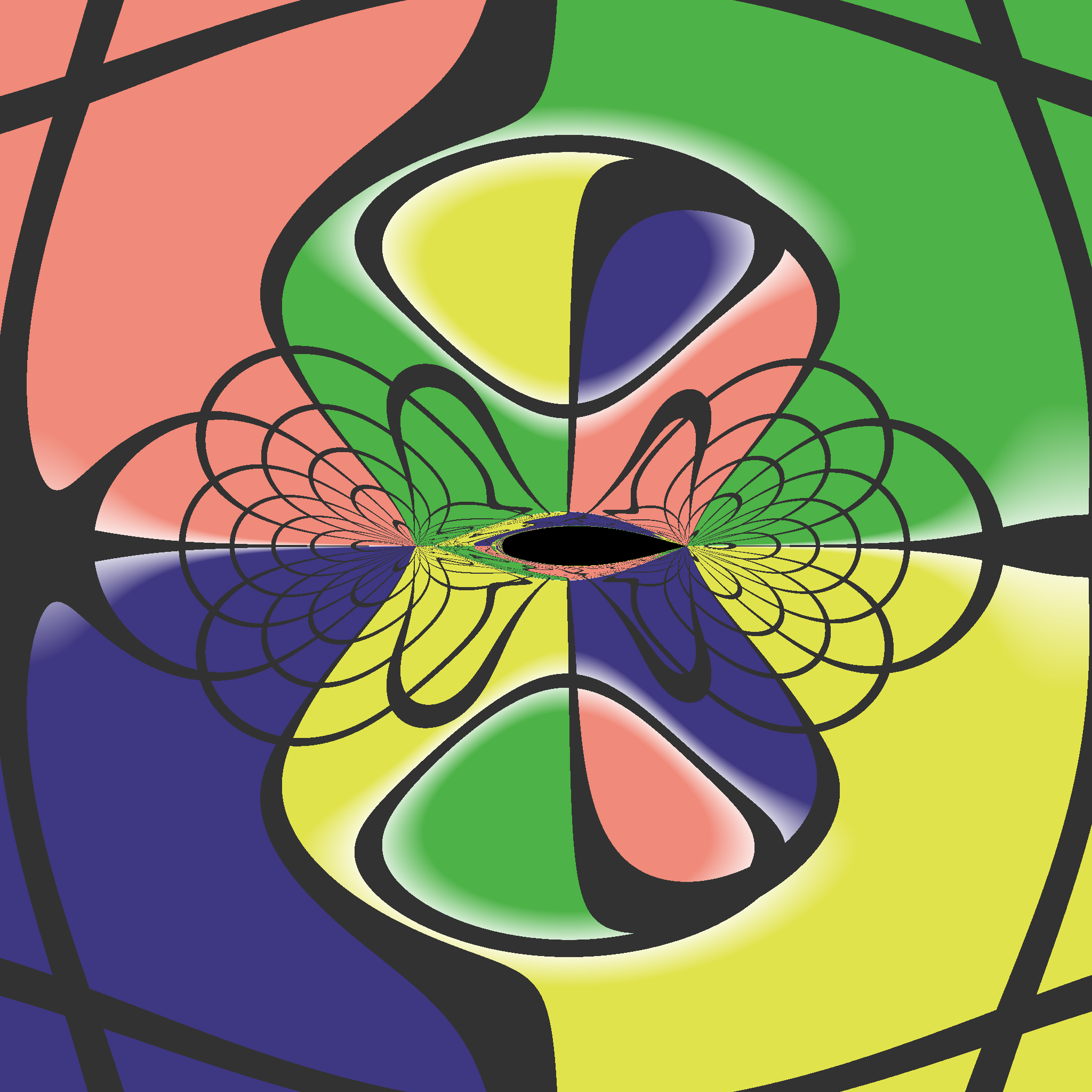}
  \end{minipage}
  }%
  \subfigure[$i=8, r_1\simeq7.2$]{
  \begin{minipage}[t]{0.19\linewidth}
  \centering
  \includegraphics[width=1.2in]{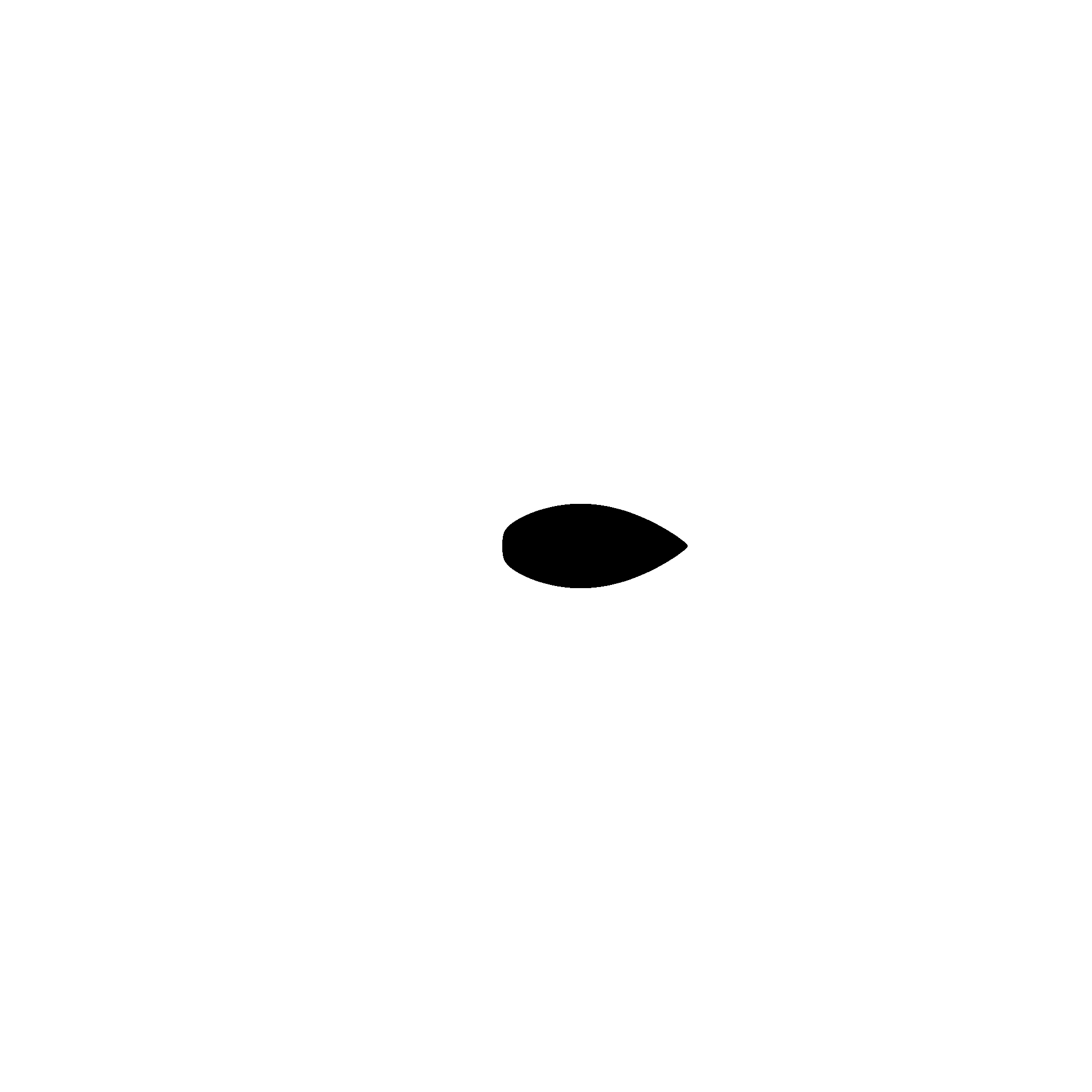}
  \end{minipage}
  }%
  \subfigure[$i=10, r_1\simeq9.0$]{
  \begin{minipage}[t]{0.19\linewidth}
  \centering
  \includegraphics[width=1.2in]{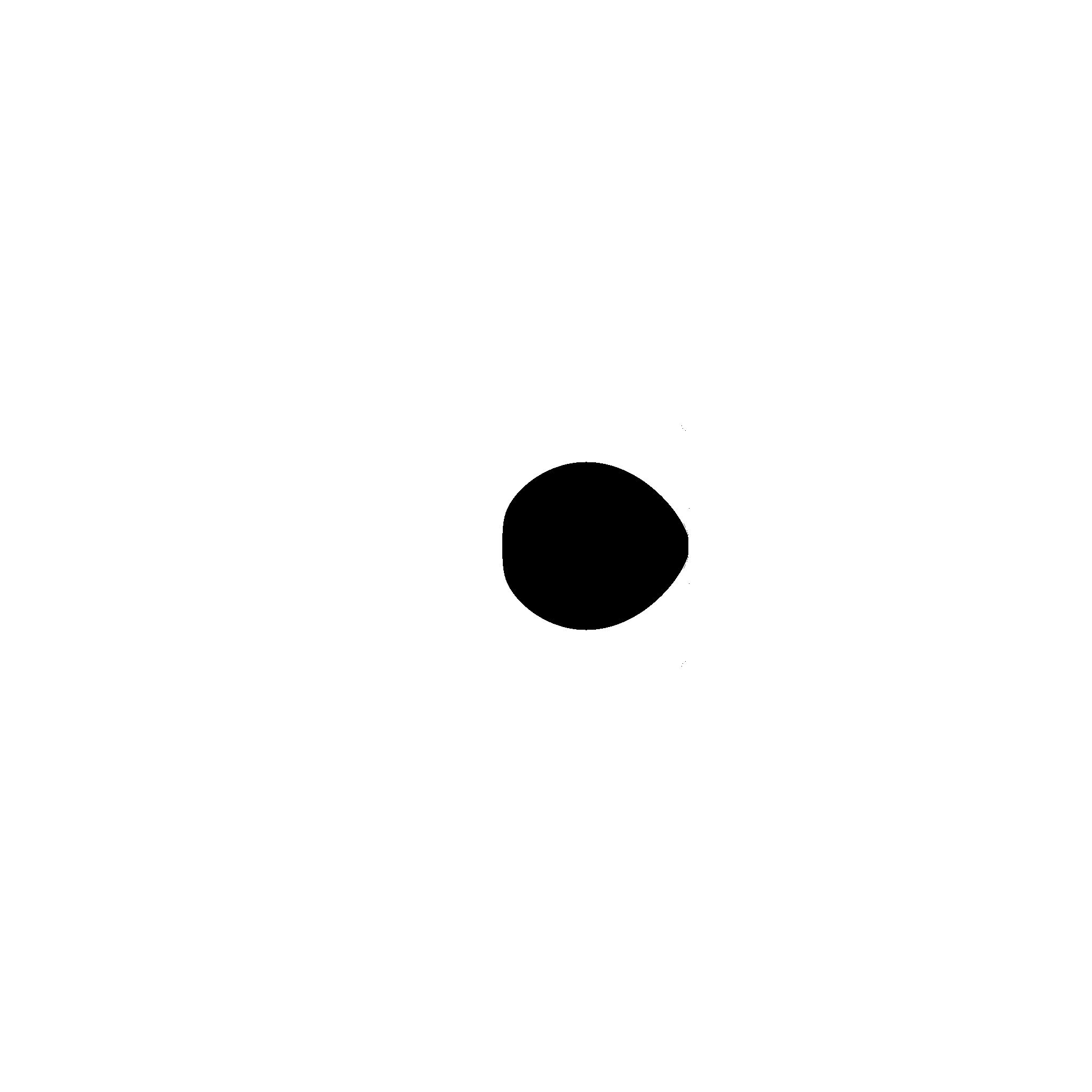}
  \end{minipage}
  }%

  \centering
  \caption{The images of the Kerr black hole surrounded by a plasma of model B for various $i$ and $r_m$. The inclination angle of the observer is $\theta_o=\pi/2$. The spin is  $a=0.998$, and $\xi_\theta=0.36, k_B=0.9$. The radial positions of maximum densities of each row from top to bottom are $2, 4, 6, 8, 10$, respectively.}
  \label{mBrmi}
\end{figure}

Now, we turn to the shadow of the Kerr black hole surrounded by the plasma of model B which is our main interest in this work. Note that the behavior with respect to $k_A$ and $k_B$ is similar in both models, so we would like to only focus on the influence of the radial position of maximum density $r_m$, the rate of the decay from the maximum towards both sides, and the opening angle $\xi_\theta$ of the plasma on the shape and size of the Kerr black hole. Recall that we have introduced the cutoff positions of the plasma as $r_1$ and $r_2$ around Eq. (\ref{dr}). Considering $r_2$ would be determined once $r_1$ is known for a given logarithmic Gaussian distribution, in order to describe the rate of the decay from the maximum, we let
\bea
r_1=\frac{r_m-r_h}{10}i\,,\quad i=1, 2, \dots, 10\,,
\eea
which indicates that a larger $i$ corresponds to a faster decay rate and a smaller width of the plasma. In particular, to include the situation that the plasma density could be nonvanishing at the event horizon, in our choice, $r_1$ could be smaller than the radius of the innermost stable circular orbit for timelike particles, which is very close to $r_h\simeq1$ when $a=0.998$, and in this case, $r_1$ is not the inner cutoff radius of the plasma when we do numerical backward ray-tracing; instead, the null trajectories terminate at the horizon.

\begin{figure}[h!]

  \centering
  \includegraphics[width=5in]{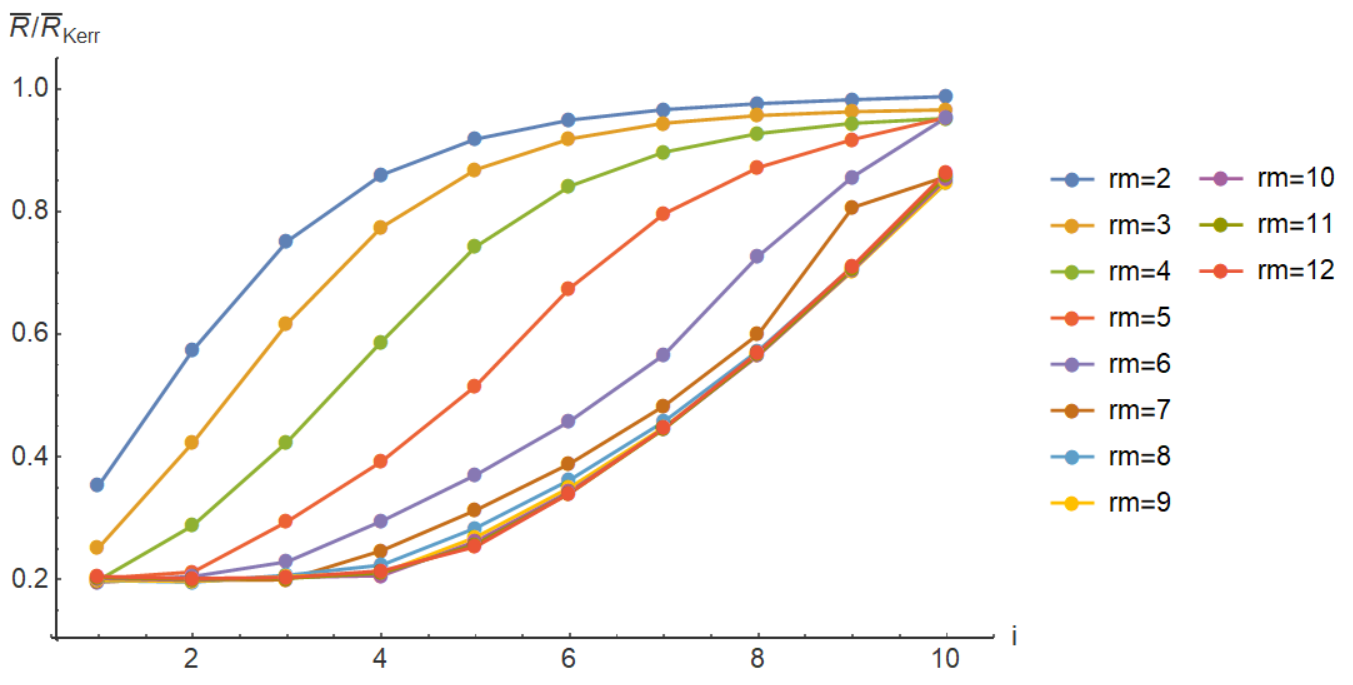}

  \caption{The variation of $\bar{R}/\bar{R}_K$ with respect to $i$ and $r_m$. The inclination angle of the observer is fixed at $\theta_o=\pi/2$. The spin is fixed at $a=0.998$ and $k_B=0.8$.}
  \label{mBda}
\end{figure}

\begin{figure}[h!]

  \centering
  \includegraphics[width=6.5in]{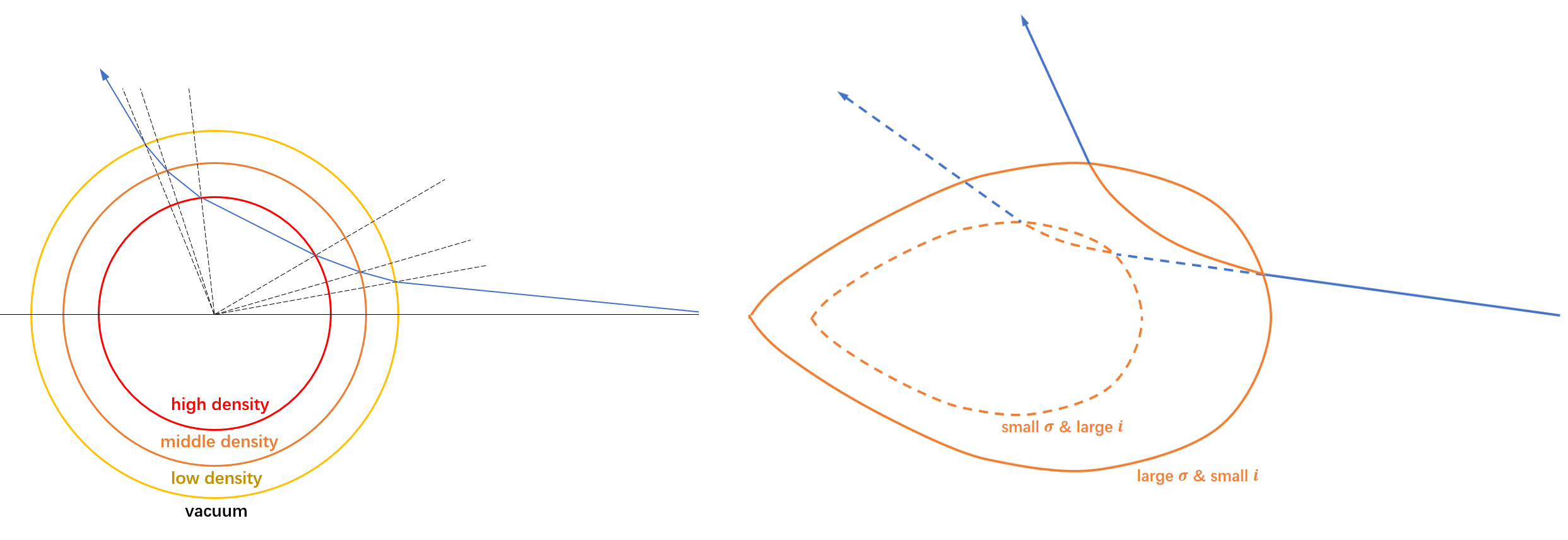}

  \caption{Two diagrams of the light trajectories traveling on the slice of $\phi=\con$. For the sake of illustration qualitatively, we use circles instead of the actual isodensity lines of the plasma in Fig. \ref{modelB} in the left plot. Since the plasma density becomes larger with the shrinking radius and the plasma's refractive index is less than 1, the trajectory diverges after multiple refractions. In the right plot, we let the two isodensity lines have the same density $N_B/N_{\text{max}}=0.1$ at a fixed $r_m$ with different $r_2$. Qualitatively, we can see that the photon would have a longer journey in the plasma of a wider plasma disk, which indicates it has a larger deflection angle when leaving the plasma.}
  \label{contour}
\end{figure}

\begin{figure}[h!]

  \centering
  \includegraphics[width=5in]{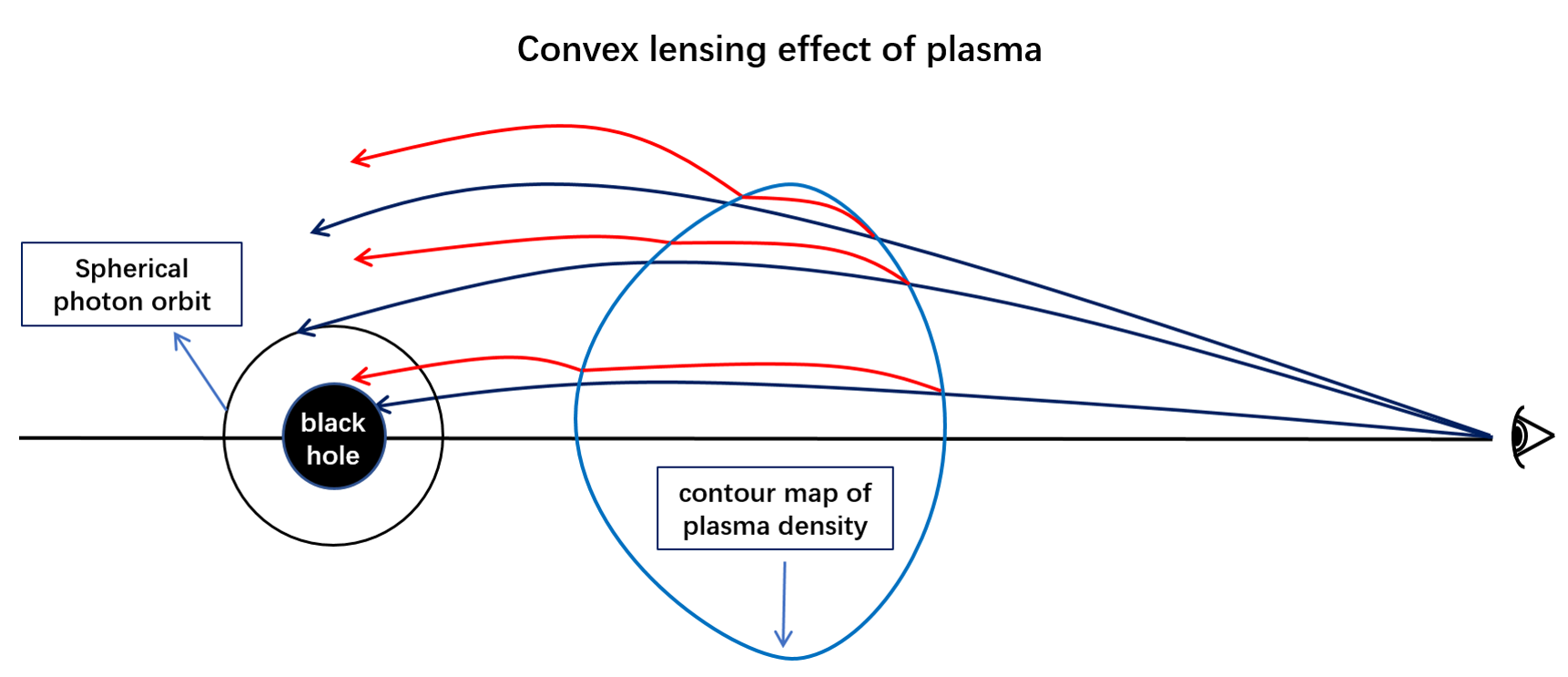}

  \caption{A diagram of the convex lensing effect of plasma on the lights.}
  \label{diapla}
\end{figure}

At first, we consider the change of the shadow curves as the maximum density position moves and  the decay rate of the density gets slower.  As mentioned at the beginning of this section, we consider the near-extreme Kerr blackhole and  fix $a=0.998$ to simplify the discussion. As a result, the radius of the innermost stable circular orbit on the equatorial plane is very near $r_h\simeq1$. In this case,   we let $r_m$ start at $2$. In Fig. \ref{mBrmi}, we show the images of the Kerr black hole surrounded by a plasma of model B with various $r_m$'s and $i$'s while the other parameters being fixed. For the first three columns, the results of gravitational lensing are included while only the shadow region is presented for the last two columns to highlight the shape of the shadow curve. From Fig. \ref{mBrmi} we can read a few remarkable features on the change of the black hole shadow. 
\begin{enumerate}
\item The size of the shadow becomes small as $i$ goes small for each row. This  means that the ability of the plasma that diverges the backward rays becomes strong as the decay rate from the maximum decreases or equivalently the width of the plasma increases.
\item The differences of the shape and size of the Kerr black hole shadow are noticeable as the increase of $r_m$ qualitatively from each column of Fig. \ref{mBrmi}. More precisely, when $r_m$ gets larger,  the size of shadow curve becomes smaller, while the change in the shape is too  complicated to read a simple rule. At different density decay rates, namely for different values of $i$, the characteristics of the shape change are different. In Fig. \ref{mBda}, we show the the variation of $\bar{R}/\bar{R}_{\mbox{Kerr}}$ with respect to $i$ and $r_m$ to quantitatively  describe the change of the black hole shadow size.
\item In particular,  for $i=8$ and $i=10$,  with the increasing of $r_m$,  there may first appear a pair of cusps in the shadow curves, which could further change into a couple of tails and disappear at last, as shown in the last two columns of Fig. \ref{mBrmi}.  \end{enumerate}
\begin{figure}[h!]
  \centering

  \subfigure[$\xi_\theta=0.09$]{
  \begin{minipage}[t]{0.3\linewidth}
  \centering
  \includegraphics[width=1.5in]{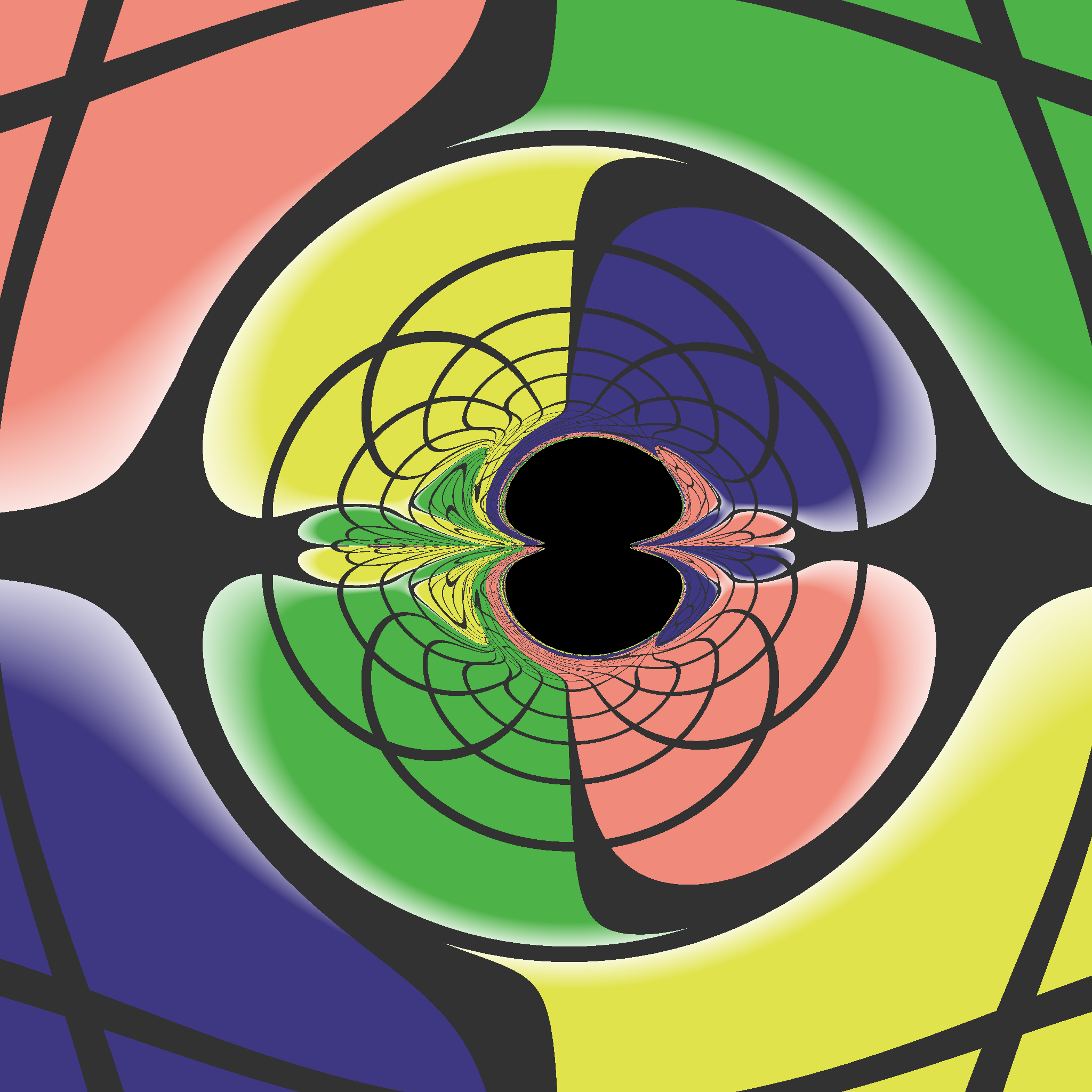}
  \end{minipage}%
  }%
  \subfigure[$\xi_\theta=0.18$]{
  \begin{minipage}[t]{0.3\linewidth}
  \centering
  \includegraphics[width=1.5in]{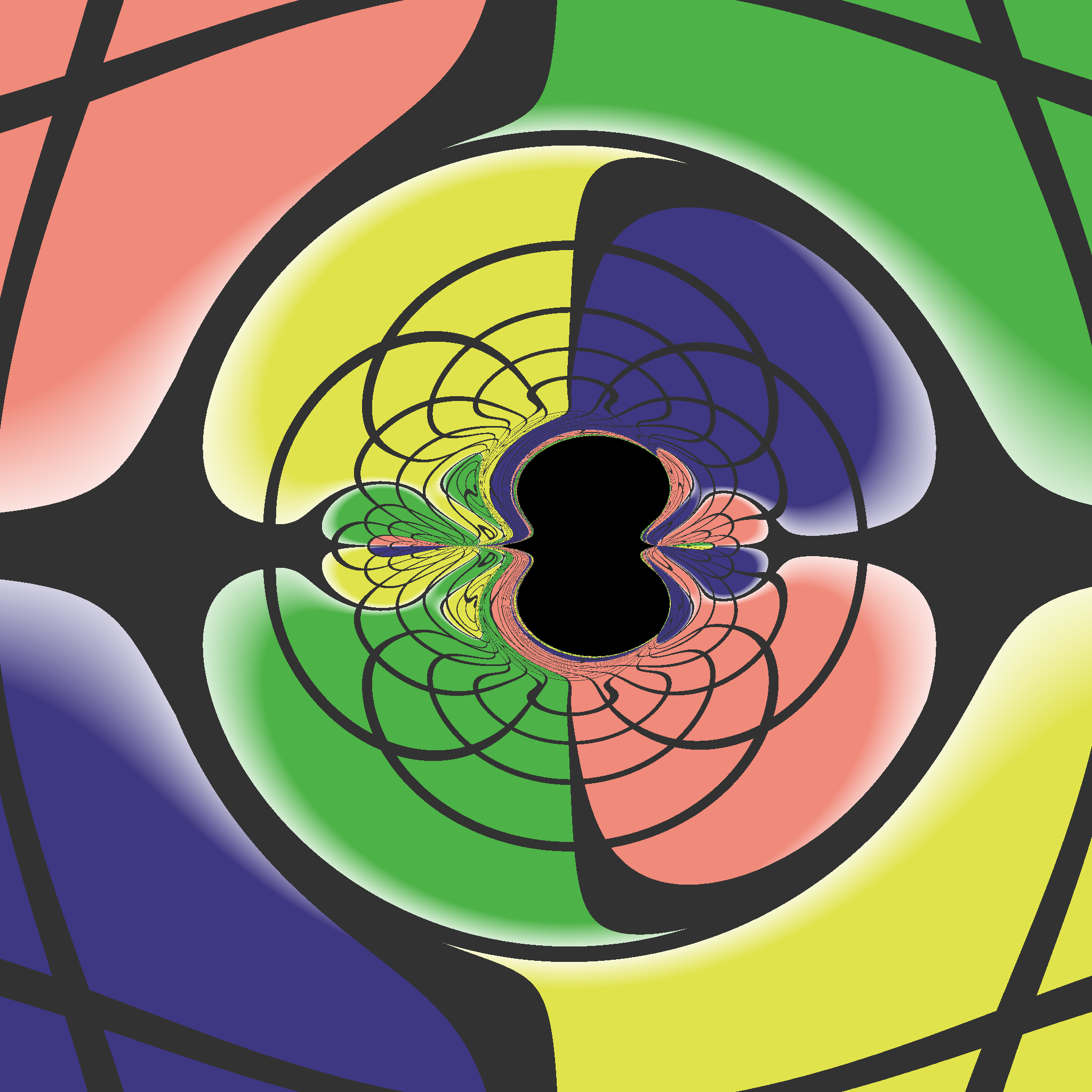}
  \end{minipage}%
  }%

  \subfigure[$\xi_\theta=0.36$]{
  \begin{minipage}[t]{0.3\linewidth}
  \centering
  \includegraphics[width=1.5in]{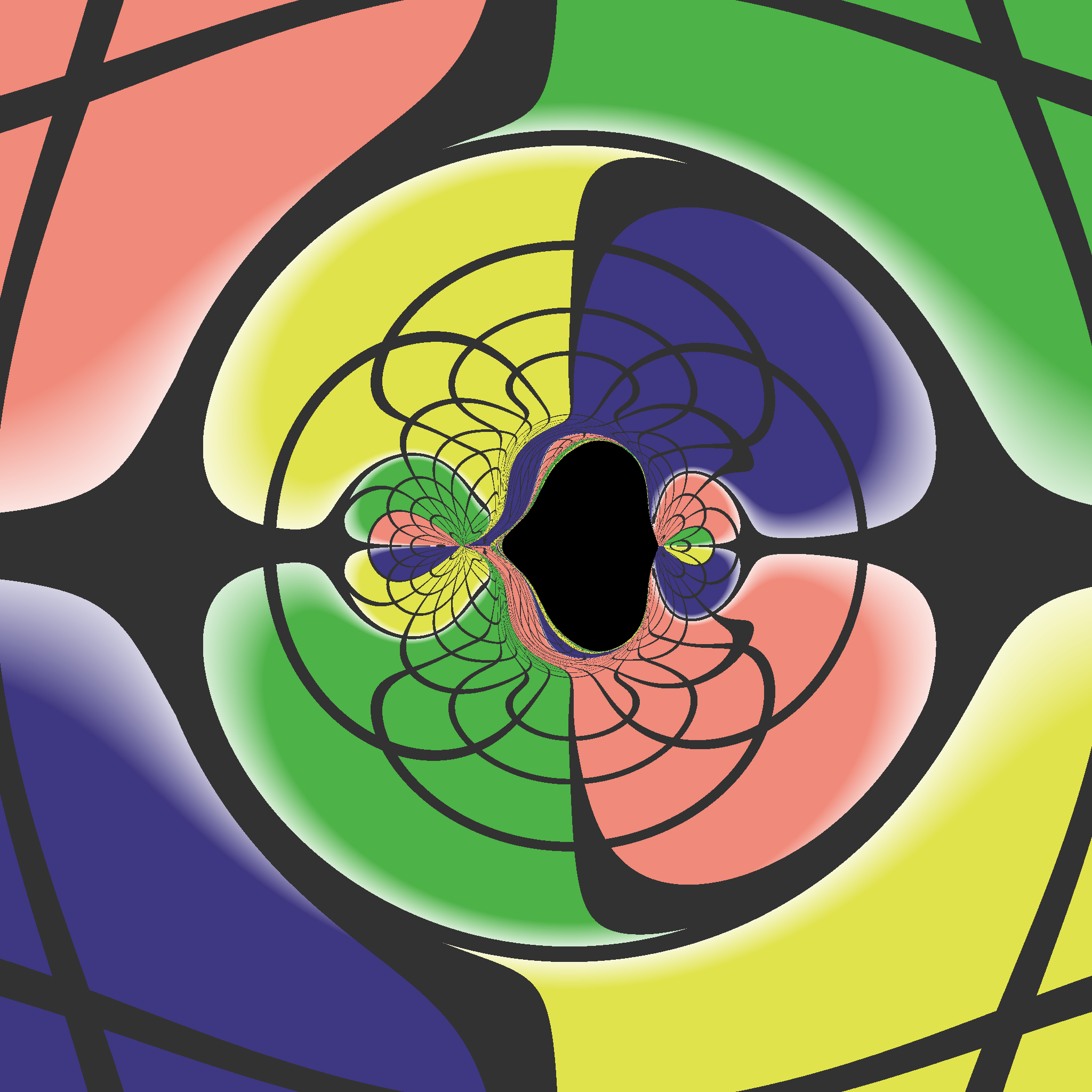}
  \end{minipage}
  }%
  \subfigure[$\xi_\theta=0.54$]{
  \begin{minipage}[t]{0.3\linewidth}
  \centering
  \includegraphics[width=1.5in]{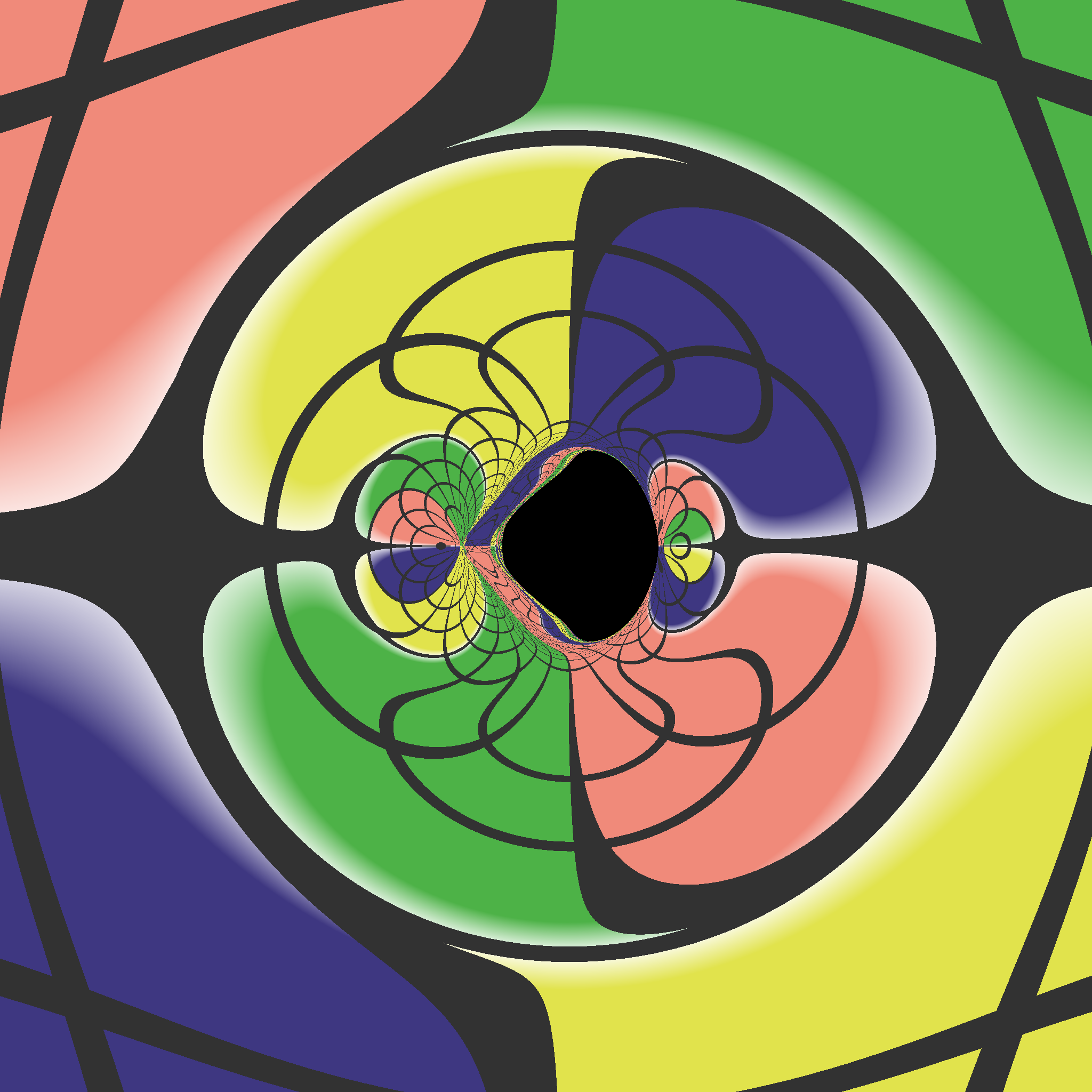}
  \end{minipage}
  }%
    \subfigure[$\xi_\theta=0.72$]{
  \begin{minipage}[t]{0.3\linewidth}
  \centering
  \includegraphics[width=1.5in]{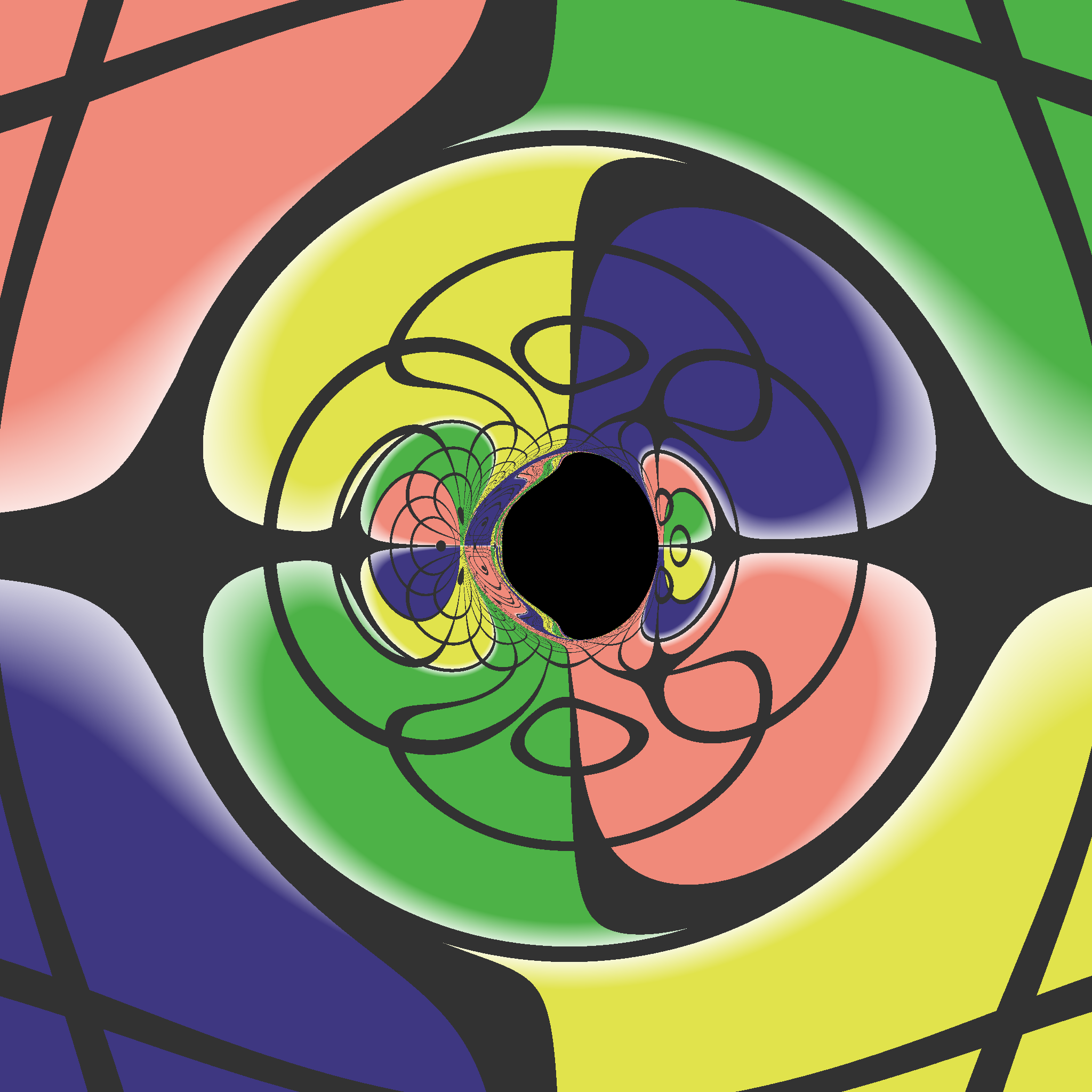}
  \end{minipage}
  }%
    
  \centering
  \caption{The images of the Kerr black hole surrounded by a plasma of model B. The inclination angle of the observer is fixed at $\theta_o=\pi/2$ and the spin is fixed at $a=0.998$. $\sigma=0.265$ and $k_B=0.8$.}
  \label{mBxi90}
\end{figure}

\begin{figure}[h!]
  \centering

\subfigure[$\xi_\theta=0.18$]{
  \begin{minipage}[t]{0.3\linewidth}
  \centering
  \includegraphics[width=1.5in]{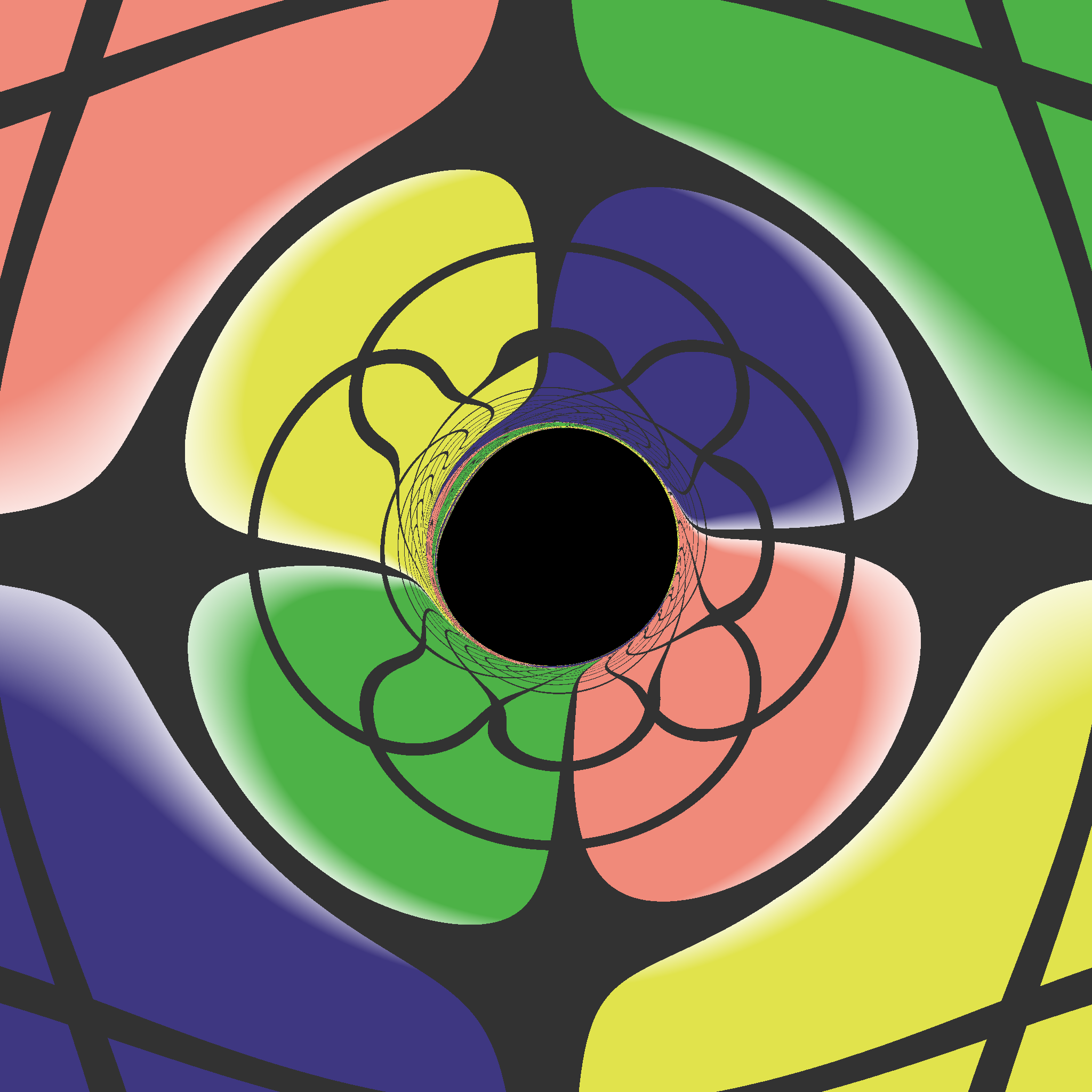}
  \end{minipage}%
  }%
  \subfigure[$\xi_\theta=0.36$]{
  \begin{minipage}[t]{0.3\linewidth}
  \centering
  \includegraphics[width=1.5in]{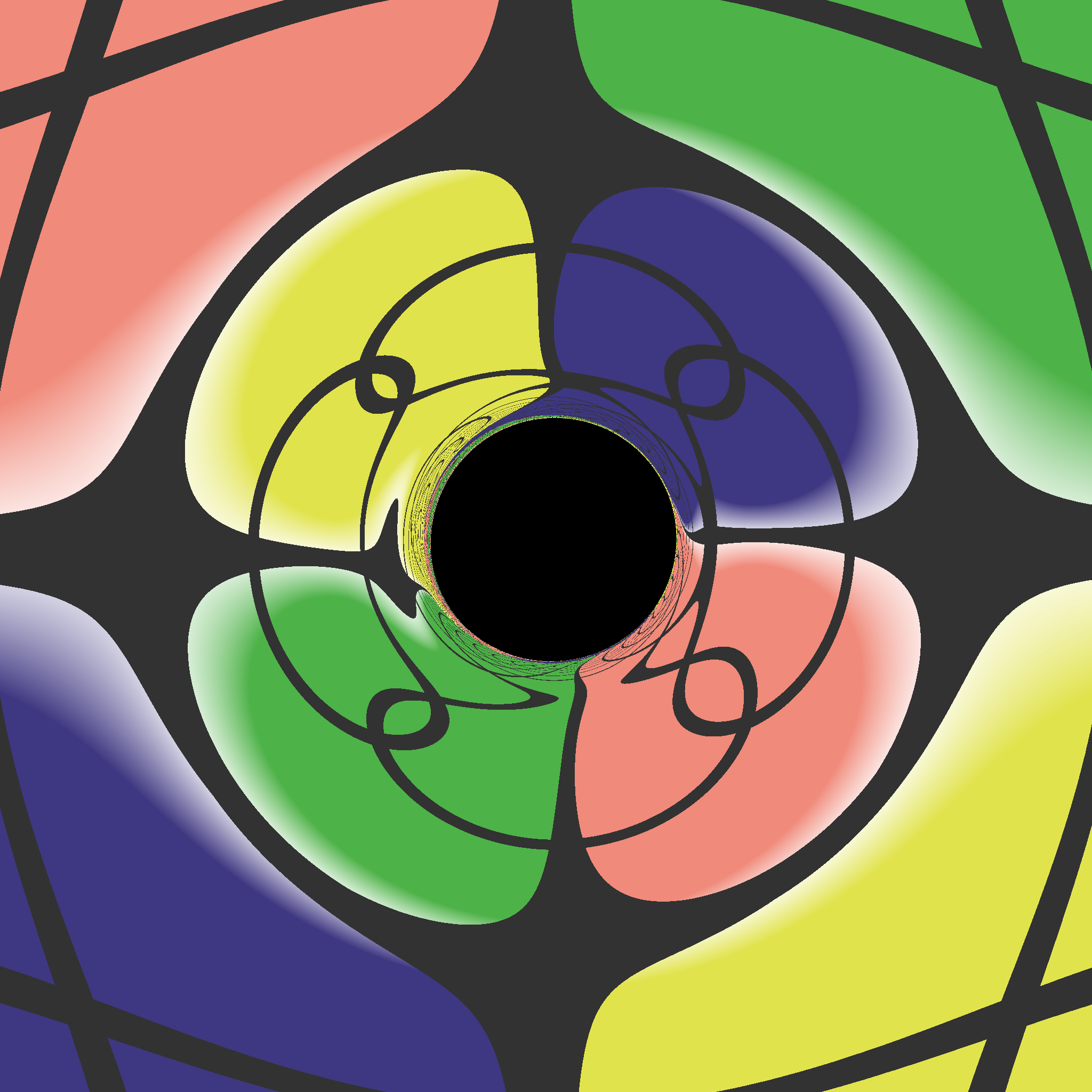}
  \end{minipage}
  }%
\subfigure[$\xi_\theta=0.72$]{
  \begin{minipage}[t]{0.3\linewidth}
  \centering
  \includegraphics[width=1.5in]{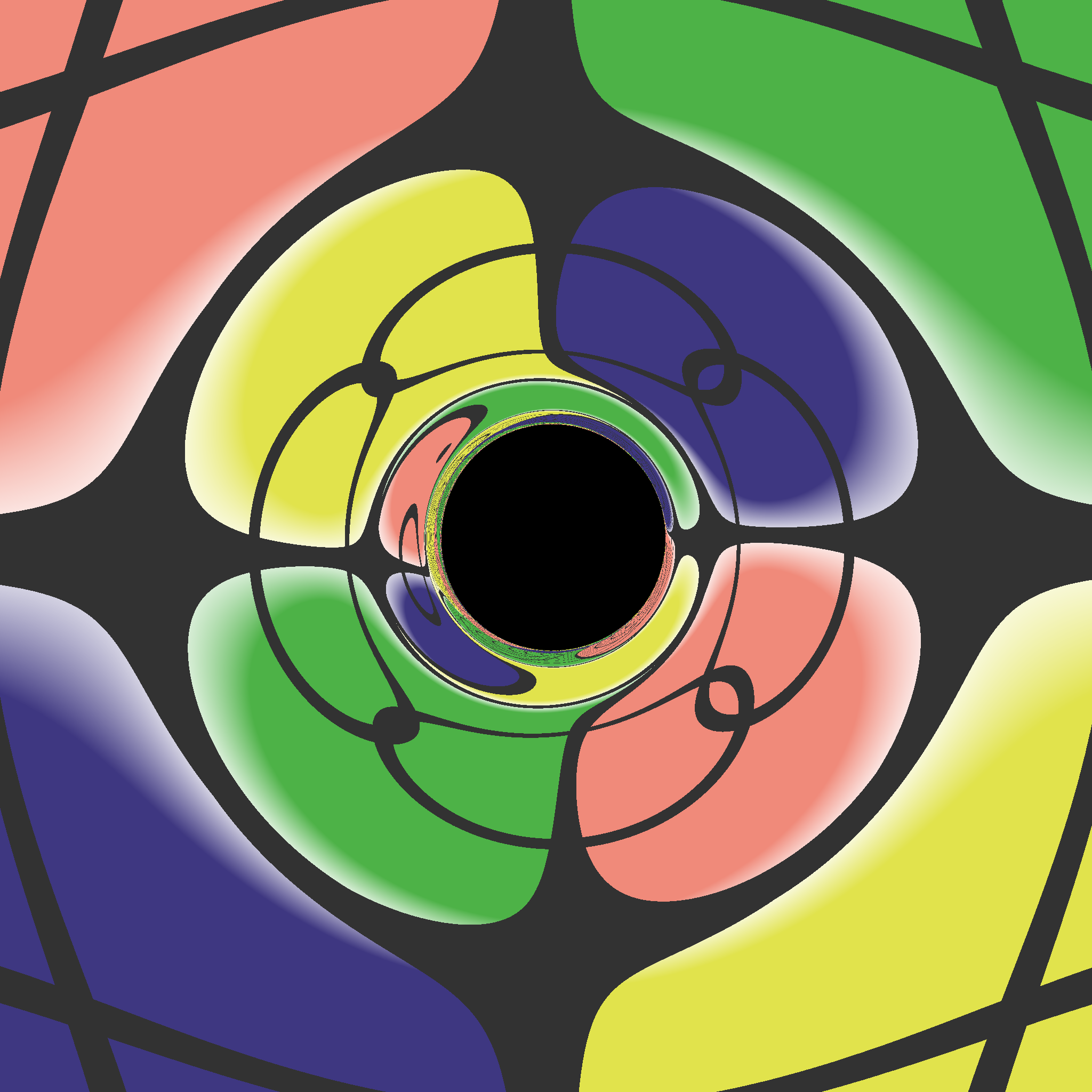}
  \end{minipage}
  }%
    
  \centering
  \caption{The images of the Kerr black hole surrounded by a plasma of model B. The inclination angle of the observer is fixed at $\theta_o=17^{\circ}$ and the spin is fixed at $a=0.998$. $\sigma=0.265$ and $k_B=0.8$.}
  \label{mBxi17}
\end{figure}

In order to illustrate the effect of plasma on light rays, we project the wave vector of the light ray to the $\phi=\con$ plane. The remaining component is perpendicular to the $\phi=\con$ plane and lies on the $z=\con$ plane. Noting that the contour map of the plasma density is the contour map of the refractive index of plasma, thus it can be regarded as an interface. On the $z=\con$ plane, the contour maps of the plasma density are circles and the component of the wave vector of the light ray is tangent to the corresponding circle. As a result, there is no deflection effect for the component of the wave vector.

The situation is different  on the two dimensional slice $\phi=\con$. As shown in Fig. \ref{modelB}, the contour maps of the plasma density are convex curves. Considering that the refractive index of plasma is less than $1$,  the closer to the center, the greater the density, and the smaller is the refractive index. In any interface,  the refraction angle should be greater than the incident angle. Therefore, such a plasma distribution will cause the light to diverge, as shown in the left plot in Fig. \ref{contour}. In addition, as we keep the maximum refractive index constant by fixing $k_B$,  a slower decay rate corresponds to a broader plasma disk. Considering the fact that photons would have longer journeys in a wider plasma, indicating a larger deflection angle when leaving the plasma, a broader plasma disk would lead to a more vital light deflection based on our model. A diagram is given in the right plot of Fig. \ref{contour}.

Roughly speaking, we could take the plasma as a convex lens in the propagation of light rays. As in Fig \ref{diapla}, there are three blue lines with arrows connecting to the observer,   each line representing a light ray in a black hole spacetime without plasma, but with different terminal points.  Among them, the top one can reach the infinity, the bottom one falls into the black hole and the middle one ends very close to the spherical photon orbit that determines a critical angle related to the black hole shadow. In the presence of a plasma of model B, the light rays would diverge after passing through the plasma, as shown in the red lines in Fig. \ref{diapla}. For example, the middle line can escape to the infinity. As a result, the size of black hole shadow would shrink. Moreover, since the increase of $i$ leads to the broadening of the plasma disk $d$,  the wider a plasma disk is, the smaller the size of the black hole shadow is.

Next, we study the influence of the opening angle $\xi_\theta$ of the plasma on the black hole shadow for model B. The results for $\theta_o=\pi/2$ and $\theta_0=17^{\circ}$ are shown in Fig. \ref{mBxi90} and Fig. \ref{mBxi17}, respectively. In addition, from the Fig. \ref{mBxi17}, we can see that the gravitational lensings change a lot as $\xi_\theta$ increases. Interestingly, compared with the results shown in Fig. \ref{mAxi} for the model A, we find that the opening angle $\xi_\theta$ has a greater influence on the shape of the black hole shadow for model B as shown in Fig. \ref{mBxi90} and Fig. \ref{mBxi17}. In particular, for $\theta_o=\pi/2$, the shape changes greatly with the increasing of the opening angle due to the large density distribution near $r_m$ for model B,  while for model A, roughly speaking, the plasma density is almost concentrated near the event horizon, as seen in Fig. \ref{modelA} and Fig. \ref{modelB}.

\section{Summary}\label{summary}
In this work, we studied the effect of plasma on the black hole shadow. We considered two models of axisymmetric plasma surrounding the Kerr black hole. Both models have an Gaussian distribution in the polar direction. Along the radial direction, the model A was assigned a distribution that the plasma density decays as power law, while for model B,  the plasma density obeyed a lognormal distribution. These plasma distributions lead to a nonseparable Hamilton-Jacobi equation, so that we employed the numerical backward ray-tracing method to investigate the effects of the plasma on the Kerr black hole shadow. We have shown some interesting and important features in the black shadow in the presence of plasma. We discussed the changes in the shape and size of the black hole shadow with different opening angles of the plasma for both models, and different radial positions of the maximum density of plasma for model B, as well as different decay rate from the maximum density toward the event horizon. 

We also observed that each contour map of the plasma density is a convex curve so that it can be seen as an interface with equal refractive index. The photon rays get diverged after passing through the plasma, considering the refractive index of plasma is less than $1$, and the closer to the center, the smaller is the refractive index on the $\phi=\con$ plane. Then, we qualitatively explained our results and concluded the existence of plasma would shrink the size of black hole shadow based on the convex lens effect of the plasma.

In addition, it is worth noting that for a possible astrophysical plasma around Sgr A* and M87*, the number density is currently believed to be $N~\sim10^{10} \text{cm}^{-3}$\cite{frank_king_raine_2002}, which corresponds to $k_A, k_B\sim10^{-4}$ (see Appendix \ref{appA} for more details). The shrinking effect caused by the plasma is weak and hard to be detected by the EHT collaborations. Nevertheless, our results are still of theoretical interest and potentially have applications in astrophysics if the systems with such high plasma densities can be found in the future.

\appendix

\section{Estimating the numerical value of the number density of the plasma}\label{appA}

In our work, we have set $c=GM=1$ for simplicity. In addition, we also let the frequency of the photon at infinity $\omega_{\infty}=1$ in the numerical calculation. Thus considering the dimension of $k_A$, we can recover the full expression $k_A^f$ as 
\bea
k_A^f=\left(\frac{GM}{c^2}\omega_\infty\right)^2k_A\,,
\eea
where $k_A$ is just a number in our paper.

On the other hand, from the relation between $\N$ and $k_A$, that is, $\N=\frac{k_A m_e}{4\pi e^2 r_h^2}$, we have
\bea
k_A^f=\frac{4\pi e^2}{m_e}\N r_h^2\,
\eea
in the Gaussian units. For a black hole, we have $r_h\sim\frac{GM}{c^2}$, thus we find 
\bea
\N=k_A \frac{m_e\omega_\infty^2}{4\pi e^2}\,.
\eea
Furthermore, the images of M87* and Sgr A* are photographed by the EHT at the observing frequency $230$ GHz, that is, $\omega_\infty=2\pi f_\infty\sim10^3 \,\text{GHz} =10^{12}\,\text{s}^{-1}$. Note that in the Gaussian units
\bea
e=4.8\times 10^{-10}\,\text{g}^{-1/2}\,\text{cm}^{3/2}\,\text{s}^{-1}\,,\quad m_e=9.1\times10^{-28}\,\text{g}\,,
\eea
then we can find
\bea
\N\sim k_A\times 10^{14}\,\text{cm}^{-3}\,.
\eea
To get an estimate of the density of the plasma, we consider the mass of hydrogen atom $\sim 10^{-24}\,\text{g}$ and find
\bea
\rho\sim k_A\times10^{-10}\,\text{g}/\text{cm}^{3}\,.
\eea
In our work, the maximum value of $k_A$ is set to be $26$, and obviously, we can see that the mass of the plasma is small enough to be neglected compared to a Kerr black hole. As a result, we are allowed to ignore the backreaction of the plasma on the background spacetime containing a Kerr black hole in our paper.

\section*{Acknowledgments}
We thank Yehui Hou, Zezhou Hu and Ye Shen  for helpful discussions. Special thanks to Haowei Sun and Zhen Zhong for their suggestions on code debugging of this work. The work is in part supported by NSFC Grant  No. 11735001, 11775022 and 11873044. MG is also supported by ``the Fundamental Research Funds for the Central Universities'' with Grant No. 2021NTST13.

\bibliographystyle{utphys}
\bibliography{plasma}

\end{document}